\documentclass[twoside]{IEEEtran}

\normalsize
\usepackage{flushend}
\usepackage{graphicx}
\usepackage{amssymb}
\usepackage{epstopdf}
\usepackage{float}
\usepackage{dblfloatfix}
\usepackage{amsmath}
\usepackage{algorithm}
\usepackage[noend]{algpseudocode}
\usepackage[justification=centering]{caption}
\usepackage{amsthm}
\usepackage{array}
\theoremstyle{remark}
\newtheorem{definition}{Definition}
\newtheorem{example}{Example}
\newtheorem{remark}{Remark}

\newtheorem{lemma}{Lemma}
\newtheorem{theorem}{Theorem}
\newtheorem{note}{Note}

\usepackage{cite}

\ifCLASSINFOpdf

\else

\fi

\title{Noisy Index Coding with PSK and QAM}
\begin{document}
	
	\author{Anjana~A.~Mahesh
		and~B.~Sundar~Rajan,~\IEEEmembership{Fellow,~IEEE}%
		\thanks{The authors are with the Department
			of Electrical Communication Engineering, Indian Institute of Science, Bengaluru 560012,
			KA, India (e-mail:~\{anjanaam,~bsrajan\}@ece.iisc.ernet.in).}
		\thanks{Part of the content of this manuscript has been accepted for publication in Proc. IEEE WCNC 2016, Doha.}
	}
	
	\markboth{}
	{Noisy Index Coding with PSK and QAM}
	
	\maketitle
	\thispagestyle{empty}	
	
	\vspace{-2cm}
	
	\begin{abstract}
		Noisy index coding problems over AWGN channel are considered. For a given index coding problem and a chosen scalar linear index code of length $N$, we propose to transmit the $N$ index coded bits as a single signal from a $2^N$- PSK constellation. By transmitting the index coded bits in this way, there is an $N/2$ - fold reduction in the bandwidth consumed. Also, receivers with side information satisfying certain conditions get coding gain relative to a receiver with no side information. This coding gain is due to proper utilization of their side information and hence is called ``PSK side information coding gain (PSK-SICG)". A necessary and sufficient condition for a receiver to get PSK-SICG is presented.  An algorithm to map the index coded bits to PSK signal set such that the PSK-SICG obtained is maximized for the receiver with maximum side information is given. We go on to show that instead of transmitting the $N$ index coded bits as a signal from $2^N$- PSK, we can as well transmit them as a signal from $2^N$- QAM and all the results including the necessary and sufficient condition to get coding gain holds. We prove that sending the index coded bits as a QAM signal is better than sending them as a PSK signal when the receivers see an effective signal set of eight points or more.
	\end{abstract}
	

	\IEEEpeerreviewmaketitle

	\section{Introduction}
	\label{sec:Intro}
	\subsection{Preliminaries}
	\label {sub:Prel}

	\IEEEPARstart
	{T}{he} noiseless index coding problem was first introduced by Birk and Kol \cite{ISCO} as an informed source coding problem over a broadcast channel. It involves a single source $\mathcal{S}$ that wishes to send $n$ messages from a set $\mathcal{X}= \lbrace x_{1},x_{2},\ldots,x_{n} \rbrace, \ x_i \in \mathbb{F}_{2}$, to a set of $m$ receivers $\mathcal{R}=\lbrace R_{1},R_{2},\ldots,R_{m} \rbrace  $. A receiver $R_{i}$ $\in$ $\mathcal{R}$ is identified by $\lbrace \mathcal{W}_i , \mathcal{K}_i \rbrace$, where $\mathcal{W}_i\subseteq \mathcal{X}$ is the set of messages demanded by the receiver $R_{i}$ and $\mathcal{K}_i\subsetneq \mathcal{X}$ is the set of messages known to the receiver $R_{i}$ a priori, called the side information set. The index coding problem can be specified by $\left(\mathcal{X},\mathcal{R}\right)$.
	
	\begin{definition}
		An index code for the index coding problem $\left(\mathcal{X},\mathcal{R}\right)$ consists of\\$1)$ An encoding function $f:\mathbb{F}_{2}^n \rightarrow \mathbb{F}_{2}^l$\\
		$2)$ A set of decoding functions $g_{1}, g_{2},\ldots,g_{m}$ such that, for a given input $\textbf{x} \in \mathbb{F}_{2}^n $, $g_{i}\left(f(\textbf{x}),\mathcal{K}_i\right) = \mathcal{W}_i, \ \forall \ i \in \lbrace 1, 2,\ldots,m \rbrace $.
	\end{definition}
	The optimal index code as defined in \cite{OMIC} is that index code which minimizes $l$, the length of the index code which is equal to the number of binary transmissions required to satisfy the demands of all the receivers. An index code is said to be linear if its encoding function is linear and linearly decodable if all its decoding functions are linear \cite{ICSI}. 
	
	The class of index coding problems where each receiver demands a single unique message were named in \cite{OMIC} as single unicast index coding problems. For such index coding problems, $m=n$. WLOG, for a single unicast index coding problem, let the receiver $R_i$ demand the message $x_i$. The side information graph $G$, of a single unicast index coding problem, is a directed graph on $n$ vertices where an edge $\left( i,j\right) $ exists if and only if $R_i$ knows the message $x_j$ \cite{ICSI}. The minrank over $\mathbb{F}_{2}$ of the side information graph $G$ is defined in \cite{ICSI} as $\min \left\lbrace rank_2\left( A\right)  : A \text{ fits } G\right\rbrace$, where a 0-1 matrix $A$ is said to fit $G$ if $a_{ii}=1 \ \forall \ i \in \left\lbrace 1,2, \ldots, n\right\rbrace$ and $a_{ij}=0$, if $\left( i,j\right)$ is not an edge in $G$ and $rank_2$ denotes the rank over $\mathbb{F}_2$. Bar-Yossef et al. in \cite{ICSI} established that single unicast index coding problems can be expressed using a side information graph and the length of an optimal index code for such an index coding problem is equal to the minrank over $\mathbb{F}_{2}$ of the corresponding side information graph.  This was extended in \cite{ECIC} to a general instance of index coding problem using minrank over $\mathbb{F}_q$ of the corresponding side information hypergraph.
	
	In both \cite{ISCO} and \cite{ICSI}, noiseless binary channels were considered and hence the problem of index coding was formulated as a scheme to reduce the number of binary transmissions. This amounts to minimum bandwidth consumption, with binary transmission. We consider noisy index coding problems over AWGN broadcast channel. Here, we can reduce bandwidth further by using some M-ary modulation scheme.  A previous work which considered index codes over Gaussian broadcast channel is by Natarajan et al.\cite {IGBC}. Index codes based on multi-dimensional QAM constellations were proposed and a metric called ``side information gain" was introduced as a measure of efficiency with which the index codes utilize receiver side information. However \cite {IGBC}  does not consider the index coding problem as originally defined in \cite{ISCO} and \cite{ICSI} as it does not minimize the number of transmissions. It always uses $2^n$- point signal sets, whereas we use signal sets of smaller sizes as well as $2^n$- point signal set for the same index coding problem.
	
	\subsection{Our Contribution}
	
	We consider index coding problems over $\mathbb{F}_2$, over AWGN broadcast channels. For a given index coding problem, for an index code of length $N$, we propose to use $2^N$- ary modulation scheme to broadcast the index codeword rather than using $N$ BPSK transmissions, with the energy of the symbol being equal to that of $N$ binary transmissions. Our contributions are summarized below.
	\begin{itemize}
		\item An algorithm, to map $N$ index coded bits to a $2^N$- PSK/ $2^N$-QAM signal set is given. 
		\item We show that by transmitting $N$ index coded bits as a signal point from $2^N$- PSK or QAM constellation, certain receivers get both coding gain as well as bandwidth gain and certain other receivers trade off coding gain for bandwidth gain.
		\item A necessary and sufficient condition that the side information possessed by a receiver should satisfy so as to get coding gain over a receiver with no side information is presented.
		\item We show that it is not always necessary to find the minimum number of binary transmissions required for a given index coding problem, i.e., a longer index code may give higher coding gains to certain receivers. 
		\item We find that for index coding problems satisfying a sufficient condition, the difference in probability of error performance between the best performing  receiver and the worst performing receiver widens monotonically with the length of the index code employed. 
		\item We prove that transmitting the $N$ index coded bits as a QAM signal is better than transmitting them as a PSK signal if the receivers see an effective signal set with eight points or more.
	\end{itemize} 
	
	\subsection{Organization}
	The rest of this paper is organized as follows. In Section \ref{sec:Model}, the index coding problem setting that we consider is formally defined with examples. The bandwidth gain and coding gain obtained by receivers by transmitting index coded bits as a PSK symbol are formally defined. A necessary and sufficient condition that the side information possessed by a receiver should satisfy so as to get coding gain over a receiver with no side information is stated and proved. In Section \ref{sec:PSK_Algo}, we give an algorithm to map the index coded bits to a $2^N$- PSK symbol such that the receiver with maximum amount of side information sees maximum PSK-SICG. In section \ref{sec:ICQAM}, we compare the transmission of index coded bits as a PSK signal against transmitting them as a QAM signal. The algorithm given in Section \ref{sec:PSK_Algo} itself can be used to map index codewords to QAM signal set. We find that all the results including the necessary and sufficient condition to get coding gain holds and show that transmitting the index coded bits as a QAM signal gives better performance if the receivers see an effective signal set with eight points or more. We go on to give examples with simulation results to support our claims in the subsequent Section \ref{sec:simu}.  Finally concluding remarks and directions for future work is given in Section \ref{sec:conc}.
	

	\section{Side Information Coding Gain}
	\label{sec:Model}
	Consider an index coding problem $\left( \mathcal{X}, \mathcal{R}\right) $, over $\mathbb{F}_2$, with $n$ messages and $m$ receivers, where each receiver demands a single message. This is sufficient since any general index coding problem can be converted into one where each receiver demands exactly one message, i.e., $\left|\mathcal{W}_i \right| = 1,\ \forall \ i \in \lbrace 1, 2, \ldots, m\rbrace$. A receiver which demands more than one message, i.e., $\left|\mathcal{W}_i \right|>1$, can be considered as $\left|\mathcal{W}_i \right|$ equivalent receivers all having the same side information set  $\mathcal{K}_i$ and demanding a single message each. Since the same message can be demanded by multiple receivers, this gives $m \geq n$.
	
	For the given index coding problem, let the length of the index code used be $N$. Then, instead of transmitting $N$ BPSK symbols, which we call the $N$- fold BPSK scheme, we will transmit a single point from a $2^{N}$- PSK signal set with the energy of the $2^{N}$- PSK symbol being equal to $N$ times the energy of a BPSK symbol, i.e., equal to the total transmitted energy of the $N$ BPSK symbols.
	
	\begin{example}
		\label{ex_psk_1} 
		Let $m=n=7$ and  $ \mathcal{W}_i = x_{i},\ \forall \ i\in \lbrace 1, 2,\ldots,7 \rbrace $. Let the side information sets be $\mathcal{K}_1 =\left\{2,3,4,5,6,7\right\},\ \mathcal{K}_2=\left\{1,3,4,5,7\right\},\ \mathcal{K}_3=\left\{1,4,6,7\right\},\ \mathcal{K}_4=\left\{2,5,6\right\},\ \mathcal{K}_5=\left\{1,2\right\},\ \mathcal{K}_6=\left\{3\right\}\ \text{and} \ \mathcal{K}_7=\phi$.\\
		The minrank over $\mathbb{F}_{2}$ of the side information graph corresponding to the above problem evaluates to $N=4$. 
		An optimal linear index code is given by the encoding matrix,
		{\small	
			\begin{center}
				$L =\left[\begin{array}{cccc}
				1 & 0 & 0 & 0\\
				1 & 0 & 0 & 0\\
				0 & 1 & 0 & 0\\
				0 & 0 & 1 & 0\\
				1 & 0 & 0 & 0\\
				0 & 1 & 0 & 0\\
				0 & 0 & 0 & 1
				\end{array}\right]$. 
			\end{center}
		}	
		The index coded bits are $\textbf{y}=\textbf{x}L$,  where,$$
		\textbf{y}= \left[y_1 \ y_2\ y_3\ y_4\right]=\left[x_1\ x_2\ \ldots x_7\right]L=\textbf{x}L$$
		
		$\mbox{giving}~~~~~~~~~~   y_{1}=x_{1}+x_{2}+x_{5};~~~~~~~~	y_{2}=x_{3}+x_{6};~~~~~~~~	y_{3}=x_{4};~~~~~~~~	y_{4}=x_7. $
	\end{example}
	
	In the 4-fold BPSK index coding scheme we will transmit 4 BPSK symbols. In the scheme that we propose, we will map the index coded bits to the signal points of a 16-PSK constellation and transmit a single complex number thereby saving bandwidth. To keep energy per bit the same, the energy of the 16-PSK symbol transmitted will be equal to the total energy of the 4 transmissions in the 4-fold BPSK scheme.
	
	
	
	
	
	This scheme of transmitting index coded bits as a single PSK signal will give bandwidth gain in addition to the gain in bandwidth obtained by going from $n$ to $N$ BPSK transmissions. This extra gain is termed as PSK bandwidth gain.
	
	\begin{definition}
		The term \textit{PSK bandwidth gain} is defined as the factor by which the bandwidth required to transmit the index code is reduced, obtained while transmitting a $2^{N}$- PSK signal point instead of transmitting $N$ BPSK signal points. 
	\end{definition}
	
	For an index coding problem, there will be a reduction in required bandwidth by a factor of $N/2$, which will be obtained by all receivers.
	
	With proper mapping of the index coded bits to PSK symbols, the algorithm for which is given in Section \ref{sec:PSK_Algo}, we will see that receivers with more amount of side information will get better performance in terms of probability of error, provided the side information available satisfies certain properties. This gain in error performance, which is solely due to the effective utilization of available side information by the proposed mapping scheme, is termed as PSK side information coding gain (PSK-SICG). Further, by sending the index coded bits as a $2^N$- PSK signal point, if a receiver gains in probability of error performance relative to a receiver in the $N$- fold BPSK transmission scheme, we say that the receiver gets PSK absolute coding gain (PSK-ACG).
	
	\begin{definition}
		The term \textit{PSK side information coding gain} is defined as the coding gain a receiver with side information gets relative to one with no side information, when the index code of length $N$ is transmitted as a signal point from a $2^{N}$- PSK constellation.
	\end{definition}
	
	\begin{definition}
		The term \textit{PSK Absolute Coding gain} is defined as the gain in probability of error performance obtained by any receiver in the $2^N$- PSK signal transmission scheme  relative to its performance in N- fold BPSK transmission scheme.
	\end{definition}
	We present a set of necessary and sufficient conditions for a receiver to get PSK-SICG in the following subsection.
	
	\subsection{PSK Side Information Coding Gain (PSK-SICG)}
	\label{subsec:PSK-SICG}
	Let $\mathcal{C} = \lbrace \textbf{y} \in \mathbb{F}_2^N \ | \ \textbf{y} = \textbf{x}L ,\ \textbf{x} \in \mathbb{F}_2^n \rbrace$, where $L$ is the $n \times N$ encoding matrix corresponding to the  linear index code chosen. Since $N \leq n $, we have $\mathcal{C} = \mathbb{F}_2^N$.
	For each of the receivers $R_{i},\ i\in \left\lbrace 1,2,\ldots, m\right\rbrace  $, define the set $S_{i}$ to be the set of all binary transmissions which $R_{i}$ knows a priori, i.e., $S_{i}= \lbrace y_{j}|y_{j}=\sum\limits_{k \in J }x_{k} ,\ J\subseteq \mathcal{K}_{i}\rbrace$. For example, in Example \ref{ex_psk_1}, $S_1= \{ y_2,y_3,y_4 \},$ $S_2= \{ y_3,y_4 \},$ $S_3= \{ y_3, y_4  \}$ and  $S_4=S_5=S_6=S_7= \phi.$ 
	
	Let $\eta_{i} = min \lbrace n-\left|\mathcal{K}_i\right|, N-\left|S_i\right|\rbrace$. For example, in Example \ref{ex_psk_1}, $\eta_1=1, \ \eta_2=\eta_3=2 \ \text{and} \ \eta_4=\eta_5=\eta_6=\eta_7=4$.

	\begin{theorem}
		\label{th_psk_1}
		A receiver $R_{i}$ will get PSK-SICG if and only if its available side information satisfies at least one of the following two conditions:
		\begin{align}
			n-\left|\mathcal{K}_i\right| < N \label{eq:Th1_1}\\
			\left|S_{i}\right|\geq 1 \label{eq:Th1_2}
		\end{align}
		
		Equivalently, a receiver $R_i$ will get PSK-SICG if and only if 
		\begin{align}
			\eta_i < N \label{eq:Th1_3}.
		\end{align}
		
		\begin{proof}
			The equivalence of the conditions in (\ref{eq:Th1_1}) and (\ref{eq:Th1_2}) and the condition in (\ref{eq:Th1_3}) is straight-forward since $\eta_i= min \lbrace n-\left|\mathcal{K}_i\right|, N-\left|S_i\right|\rbrace$ will be less than $N$ if and only if at least one of the two conditions given in (\ref{eq:Th1_1}) and (\ref{eq:Th1_2}) is satisfied. 
			
			Let $\mathcal{K}_i = \lbrace i_{1}, i_2, \ldots, i_{\left|\mathcal{K}_i\right|} \rbrace$ and $\mathcal{A}_i$ $\triangleq$ $\mathbb{F}_2^{\left|\mathcal{K}_i\right|}$, $i=1,2,\ldots,m$.
			
			\textit{Proof of the "if part"} : If condition (\ref{eq:Th1_1}) is satisfied, the ML decoder at $R_i$ need not search through all codewords in $\mathcal{C}$. For a given realization of $( x_{i_1},x_{i_2}, \ldots, x_{i_{\left|\mathcal{K}_i\right|}} )$, say, $(a_{1},a_{2}, \ldots, a_{\left|\mathcal{K}_i\right|}) \in \mathcal{A}_i$, the decoder needs to search through only the codewords in $$\left\lbrace \mathbf{y}=\mathbf{x}L : ( x_{i_1},x_{i_2}, \ldots, x_{i_{\left|\mathcal{K}_i\right|}} ) = (a_{1},a_{2}, \ldots, a_{\left|\mathcal{K}_i\right|})\right\rbrace, $$ i.e., the codewords in $\mathcal{C}$ which resulted from $\mathbf{x}$ such that $( x_{i_1},x_{i_2}, \ldots, x_{i_{\left|\mathcal{K}_i\right|}} ) = (a_{1},a_{2}, \ldots, a_{\left|\mathcal{K}_i\right|})$. Since number of such $\mathbf{x}$ is $= 2^{n-\left|\mathcal{K}_i\right|} < 2^N$, the decoder need not search through all the codewords in $\mathcal{C}$. 
			
			Similarly if the condition (\ref{eq:Th1_2}) is satisfied, then also the ML decoder at $R_i$ need not search through all the codewords in $\mathcal{C}$. For any fixed realization of $( x_{i_1},x_{i_2}, \ldots, x_{i_{\left|\mathcal{K}_i\right|}} )$, the values of $\left\lbrace y_j \in S_i\right\rbrace $ are also fixed. The decoder needs to search through only those $\mathbf{y} \in \mathcal{C}$ with the given fixed values of $\left\lbrace y_j \in S_i\right\rbrace $. Again, the number of such  $\mathbf{y}$ is less than $2^N$. 
			
			Thus, if any of the two conditions of the theorem is satisfied, the ML decoder at $R_i$ need to search through a reduced number of signal points, which we call the effective signal set seen by $R_i$. The size of the effective signal set seen by the receiver is $2^{\eta_i} < 2^N$. Therefore, by appropriate mapping of the index coded bits to PSK symbols, we can increase $d_{min}(R_i) \triangleq$ the minimum distance of the effective signal set seen by the receiver $R_i$, $i = 1,2,\ldots,m$, thus getting PSK-SICG.

			\textit{Proof of the "only if part"} : If none of the two conditions of the theorem are satisfied or equivalently if $\eta_i \nless N$, then the effective signal set seen by $R_i$ will be the entire $2^N$-PSK signal set. Thus $d_{min}(R_i)$ cannot be increased. $d_{min}(R_i)$ will remain equal to the minimum distance of the  corresponding $2^{N}$-- PSK signal set. Therefore the receiver $R_i$ will not get PSK-SICG.
		\end{proof}
	\end{theorem}
	
	\begin{note}
		The condition (\ref{eq:Th1_2}) above indicates how the PSK side information coding gain is influenced by the linear index code chosen. Different index codes for the same index coding problem will give different values of $\left|S_{i}\right|, \ i \in \left[ m\right] $ and hence leading to possibly different PSK side information coding gains.
	\end{note}
	\begin{figure*}
		\includegraphics[scale=0.28]{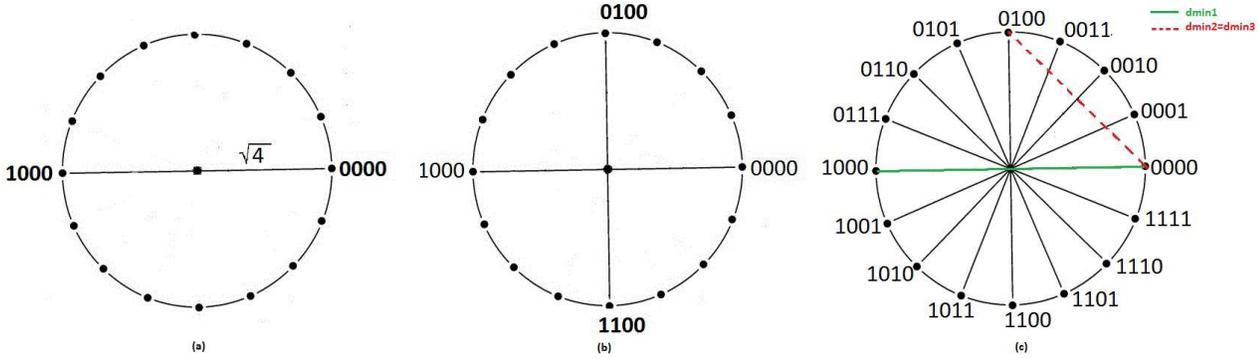}
		\caption{16-PSK Mapping for Example \ref{ex_psk_1}.}
		\label{fig:ex1_map}
		~\hrule
	\end{figure*}
	Consider the receiver $R_1$ in Example \ref{ex_psk_1}. It satisfies both the conditions with $n-\left|\mathcal{K}_1\right|  = 7-6 = 1 < 4$ and $\left|S_{1}\right| = 3 >1 $. For a particular message realization $(x_1, x_2, \dots, x_7)$, the only index coded bit $R_1$ does not know a priori is $y_1$. Hence there are only 2 possibilities for the received codeword at the receiver $R_1$. Hence it needs to decode to one of these 2 codewords, not to one of the 16 codewords that are possible had it not known any of $y_1,\ y_2,\ y_3,\ y_4$ a priori. Then we say that $R_1$ sees an effective codebook  of size 2. This reduction in the size of the effective codebook seen by the receiver $R_1$ is due to the presence of side information that satisfied condition (\ref{eq:Th1_1}) and (\ref{eq:Th1_2}) above. 
	
	For a receiver to see an effective codebook of size $ < 2^N $, it is not necessary that the available side information should satisfy both the conditions. If at least one of the two conditions is satisfied, then that receiver will see an effective codebook of reduced size and hence will get PSK-SICG by proper mapping of index coded bits to $2^N$- PSK symbols. This can be seen from the following example.
	
	\begin{example}
		\label{ex3}
		Let $m=n=6$ and $\mathcal{W}_i = x_{i}, \ \forall \ i\in \lbrace 1, 2,\ldots,6 \rbrace $. Let the known sets be $\mathcal{K}_1 =\left\{2,3,4,5,6\right\},\ \mathcal{K}_2=\left\{1,3,4,5\right\},\ \mathcal{K}_3=\left\{2,4,6\right\},\ \mathcal{K}_4=\left\{1,6\right\},\ \mathcal{K}_5=\left\{3\right\} \ \text{and} \ \mathcal{K}_6=\phi$.\\
		The minrank over $\mathbb{F}_{2}$ of the side information graph corresponding to the above problem evaluates to $N=4$. 
		An optimal linear index code is given by the encoding matrix,
		{\small	
			\begin{center}		
				$L = \left[\begin{array}{cccc}
				1 & 0 & 0 & 0\\
				0 & 1 & 0 & 0\\
				0 & 1 & 0 & 0\\
				1 & 0 & 0 & 0\\
				0 & 0 & 1 & 0\\
				0 & 0 & 0 & 1
				\end{array}\right]$ \\
				
			\end{center}
		}	
		The index coded bits in this example are,
		$$ y_{1}=x_{1}+x_{4};~~~~~~~~ y_{2}=x_{2}+x_{3};~~~~~~~~ y_{3}=x_{5}; ~~~~~~~~	y_{4}=x_{6}. $$
	\end{example}
	Here, receiver $R_4$ does not satisfy condition (1) since $n-\left|\mathcal{K}_4\right|  = 6-2 = 4 = N$.
	However, it will still see an effective codebook of size 8, since  $\left|S_{4}\right| = 1 $, and hence will get PSK-SICG by proper mapping of the codewords to 16-PSK signal points.
	\begin{note}
		The condition required for a receiver $R_i$ to get PSK-ACG is that the minimum distance of the effective signal set seen by it, $d_{min}(R_i) > 2$ since the minimum distance seen by any receiver while using $N$-fold BPSK to transmit the index coded bits is $d_{min}(\text{BPSK})=2$.
	\end{note}
	
	\begin{note}
		For the class of index coding problems with $\mathcal{W}_i \cap \mathcal{W}_j = \phi, \ \mathcal{K}_i \cap \mathcal{K}_j = \phi , \ i \neq j$ and $\left| \mathcal{W}_i\right|  = 1, \ \left| \mathcal{K}_i\right|  = 1$, which were called single unicast single uniprior in \cite{OMIC}, $\left|\mathcal{S}_i\right| = 0, \ \forall \ i \in \lbrace 1,2, \ldots, m \rbrace$. Therefore, no receiver will get PSK-SICG.
	\end{note}
	
	\subsection{$2^N$-PSK to $2^n$-PSK}
	In this subsection we discuss the effect of the length of the index code used on the probability of error performance of different receivers. We consider index codes of all lengths from the minimum length $N=$ minrank over $\mathbb{F}_2$ of the corresponding side information hypergraph  to the maximum possible value of $N=n$. Consider the following example. 
	%
	\begin{example}
		\label{ex_N_to_n1}
		Let $m=n=5$ and  $\mathcal{W}_i = \lbrace x_i \rbrace, \ \forall \ i \in \lbrace 1, 2, 3, 4, 5 \rbrace.$ Let the known information be $ \mathcal{K}_1=\lbrace2,3,4,5\rbrace, \ \mathcal{K}_2=\lbrace1,3,5\rbrace, \ \mathcal{K}_3=\lbrace1,4\rbrace, \ \mathcal{K}_4= \lbrace2\rbrace$ and $ \mathcal{K}_5 = \phi$.
		
		For this problem, minrank, $N$ = 3. An optimal linear index code is given by 
		{\small
			\begin{align*}
				L_1 = \left[\begin{array}{ccc}
					1 & 0 & 0\\
					1 & 1 & 0\\
					1 & 0 & 0\\
					0 & 1 & 0\\
					0 & 0 & 1
				\end{array}\right],
			\end{align*}
		}
		with the index coded bits being
		$$ y_{1}=x_{1}+x_{2}+x_{3};~~~~~ y_{2}=x_{2}+x_{4}; ~~~~~ y_{3}=x_{5}. 
		$$
		%
		%
		Now, we consider an index code of length $N+1=4$. The corresponding encoding matrix is 
		{\small
			\begin{align*}
				L_2 = \left[\begin{array}{cccc}
					1 & 0 & 0 & 0\\
					1 & 0 & 0 & 0\\
					0 & 1 & 0 & 0\\
					0 & 0 & 1 & 0\\
					0 & 0 & 0 & 1
				\end{array}\right]
			\end{align*}
		}
		and the index coded bits are
		$$ y_{1}=x_{1}+x_{2};~~~~~ y_{2}=x_{3};~~~~ y_{3}=x_{4};~~~~ y_{4}=x_{5}.
		$$
		We compare these with the case where we send the messages as they are, i.e., 
		\begin{align*}
			L_3= I_5, 
		\end{align*}
		where $I_5$ denotes the $5 \times 5$ identity matrix.
		Optimal mappings for the three different cases considered are given in Fig. \ref{fig:ex_N_to_n1}(a), (b) and (c) respectively.
		
		\begin{figure*}
			\begin{center}		
				\includegraphics[scale=0.4]{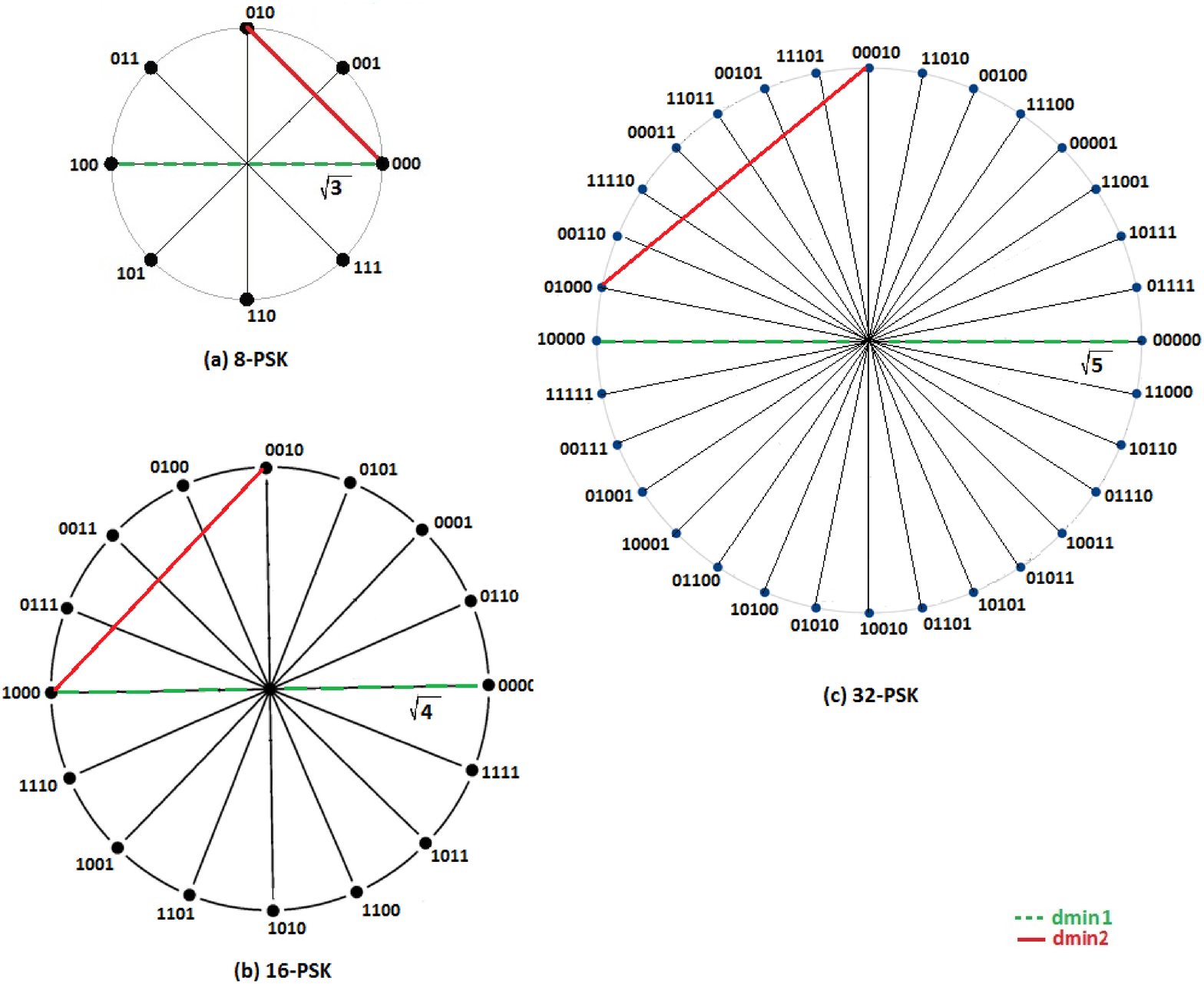}
				\caption{8-PSK, 16-PSK and 32-PSK Mappings for  Example \ref{ex_N_to_n1}.}
				\label{fig:ex_N_to_n1}
			\end{center}
			~\hrule
		\end{figure*}
		The values of $\eta$ for the different receivers while using the three different index codes are summarized in TABLE \ref{table_eta}. We see that the receiver $R_1$ sees a two point signal set irrespective of the length of the index code used. Since as the length of the index code increases, the energy of the signal also increases, $R_1$ will see a larger minimum distance when a longer index coded is used. However, the minimum distance seen by the receiver $R_5$ is that of the $2^N$ signal set in all the three cases, which decreases as $N$ increases. Hence the difference between the performances of $R_1$ and $R_5$ increases as the length of the index code increases. This is generalized in the following lemma. 
		\begin{table}
			\begin{center}
				\begin{tabular}{|c|c|c|c|c|c|}
					\hline $N$ & $\eta_1$ & $\eta_2$ & $\eta_3$ & $\eta_4$ & $\eta_5$ \\ 
					\hline 3 & 1 & 2 & 3 & 3 & 3 \\ 
					4 & 1 & 2 & 3 & 4 & 4 \\ 
					5 & 1 & 2 & 3 & 4 & 5 \\ 
					\hline 
				\end{tabular} 
				\caption{Table showing values of $\eta$ for different receivers.}
				\label{table_eta}
			\end{center}	
		\end{table}
		\begin{lemma}
			\label{lemm:N_to_n}
			For a given index coding problem, as the length of the index code used increases from N to n, where N is the minrank of the side information hypergraph of the index coding problem and n is the number of messages, the difference in performance between the best performing and the worst performing receiver increases monotonically if the worst performing receiver has no side information, provided we use an optimal mapping of index coded bits to PSK symbols given by Algorithm \ref{algo_psk}.
			\begin{proof}
				If there is a receiver with no side information, say $R$, whatever the length $l$ of the index code used, the effective signal set seen by $R$ will be $2^l$- PSK. Therefore the minimum distance seen by $R$ will be the minimum distance of $2^l$- PSK signal set. For PSK symbol energy $l$, the squared minimum pair-wise distance of $2^l$- PSK, $d_{min}(2^l$- PSK$)$, is given by  $d_{min}(2^l$- PSK$) = 4lsin^{2}(\pi /(2^l))$, which is monotonically decreasing in $l$.
			\end{proof}
		\end{lemma}
		
		\begin{remark}
			\label{rem:N_to_n}
			For an index coding problem where the worst performing receiver knows one or more messages a priori, whether or not the gap between the best performing receiver and the worst performing receivers widens monotonically depends on the index code chosen. This is because the index code chosen determines $\eta$ of the receivers which in turn determines the mapping scheme and thus the effective signal set seen by the receivers. Therefore the minimum distance seen by the receivers and thus their error performance depends on the index code chosen.
		\end{remark}
		\section{Algorithm}
		\label{sec:PSK_Algo}
		In this section we present the algorithm for labelling the appropriate sized PSK signal set. Let the number of binary transmissions required $=$ minrank over $\mathbb{F}_{2}=N$ and the $N$ transmissions are labeled $Y= \lbrace y_{1},y_{2},\ldots,y_{N} \rbrace$, where each of $y_{i}$ is a linear combination of $\lbrace x_{1},x_{2},\ldots,x_{n} \rbrace$. If the minrank is not known then $N$ can be taken to be the length of any known linear index code. 
		
		Order the receivers in the non-decreasing order of $\eta_{i}$.
		WLOG, let $\lbrace R_{1},R_{2},\ldots,R_{m} \rbrace$ be such that 
		\begin{align*}
			\eta_{1} \leq \eta_2 \leq \ldots \leq \eta_m.
		\end{align*}
		
		Let $\mathcal{K}_i = \lbrace i_{1}, i_2, \ldots, i_{\left|\mathcal{K}_i\right|} \rbrace$ and $\mathcal{A}_i$ $\triangleq$ $\mathbb{F}_2^{\left|\mathcal{K}_i\right|}$, $i=1,2,\ldots,m$. As observed in the proof of Theorem \ref{th_psk_1}, for any given realization of $( x_{i_1},x_{i_2}, \ldots, x_{i_{\left|\mathcal{K}_i\right|}} )$, the effective signal set seen by the receiver $R_i$ consists of $2^{\eta_i}$ points. Hence if $\eta_{i} \geq N $, then $d_{min}(R_i) =$ the minimum distance of the signal set seen by the receiver $R_i$, $i = 1,2,\ldots,m$, will not increase. $d_{min}(R_i)$ will remain equal to the minimum distance of the  corresponding $2^{N}$- PSK. Thus for receiver $R_i$ to get PSK-SICG, $\eta_{i}$ should be less than $ N $.

		The algorithm to map the index coded bits to PSK symbols is given in \textbf{Algorithm 1}.
		
		Before running the algorithm, Use Ungerboeck set partitioning \cite{TCM} to partition the $2^{N}$- PSK signal set into $N$ different layers. Let $L_0,\ L_1, ..., L_{N-1}$ denote the different levels of partitions of the $2^N$-PSK with the minimum distance at layer $L_i=\Delta_i$, $i \in \lbrace0, 1,\ldots, N-1\rbrace$, being such that $\Delta_0 < \Delta_1 <  \ldots < \Delta_{N-1}$.

		\begin{algorithm}
			\caption{Algorithm to map index coded bits to PSK symbols}\label{algo_psk}
			\begin{algorithmic}[1]
				
				\If {$\eta_1 \geq N $}, do an arbitrary order mapping and \textbf{exit}.
				\EndIf
				
				\State $i \gets 1$
				
				\If {all $2^N$ codewords have been mapped}, \textbf{exit}.
				\EndIf
				
				\State  Fix $( x_{i_1},x_{i_2}, \ldots, x_{i_{\left|\mathcal{K}_i\right|}} )=(a_{1},a_{2}, \ldots, a_{\left|\mathcal{K}_i\right|}) \in \mathcal{A}_i$ such that the set of codewords, $\mathcal{C}_i \subset \mathcal{C} $, obtained by running all possible combinations of $\lbrace x_{j}|\ j \notin \mathcal{K}_i\rbrace$ with $( x_{i_1},x_{i_2}, \ldots, x_{i_{\left|\mathcal{K}_i\right|}} )=(a_{1},a_{2}, \ldots, a_{\left|\mathcal{K}_i\right|})$ has maximum overlap with the codewords already mapped to PSK signal points.
				
				\If {all codewords in $\mathcal{C}_i$ have been mapped},
				\begin{itemize}
					\item $\mathcal{A}_i$=$\mathcal{A}_i \setminus \lbrace( x_{i_1},x_{i_2}, \ldots, x_{i_{\left|\mathcal{K}_i\right|}} )|( x_{i_1},x_{i_2}, \ldots, x_{i_{\left|\mathcal{K}_i\right|}} )$ together with all combinations of $\lbrace x_{j}|\ j \notin \mathcal{K}_i\rbrace$ will result in $\mathcal{C}_i\rbrace$.
					\item $i \gets i+1$
					\item \textbf{if} {$\eta_i \geq N$} \textbf{then},
					\begin{itemize}
						\item $i \gets 1$.
						\item goto \textbf{Step 3}
					\end{itemize} 
					\item \textbf{else}, goto \textbf{Step 3}
					
				\end{itemize}

				\Else
				\begin{itemize}
					\item Of the codewords in $\mathcal{C}_i$ which are yet to be mapped, pick any one and map it to a PSK signal point in that $2^{\eta_i}$ sized subset at level $L_{N-\eta_i}$ which has maximum number of signal points mapped by codewords in $\mathcal{C}_i$ without changing the already labeled signal points in that subset. 			
					If all the signal points in such a subset have been already labeled, then map it to a signal point in another $2^{\eta_i}$ sized subset at the same level $L_{N-\eta_i}$  such that this point together with the signal points corresponding to already mapped codewords in $\mathcal{C}_i$ has the largest minimum distance possible. Clearly this minimum distance, $d_{min}(R_i)$ is such that $\Delta_{N-\eta_i} \geq d_{min}(R_i) \geq \Delta_{N-(\eta_i+1)}$.
					\item $i \gets 1$
					\item goto \textbf{Step 3}
				\end{itemize}
				\EndIf

			\end{algorithmic}
		\end{algorithm}
		
		Let the PSK-SICG obtained by the mapping given in Algorithm \ref{algo_psk} by the receiver $R_i = g_i, \ i \in \left\lbrace 1,2,\ldots, m\right\rbrace $. This algorithm  gives an optimal mapping of index coded bits to PSK symbols. Here optimality is in the sense that, for the receivers  $\lbrace R_{1},R_{2},\ldots,R_{m} \rbrace$ ordered such that 
		$\eta_{1} \leq \eta_2 \leq \ldots \leq \eta_m$, 
		\begin{enumerate}
			\item No other mapping can give a PSK-SICG $> g_1$ for the receiver $R_1$.
			\item Any mapping which gives PSK-SICG $=g_j$ for the receivers $R_j, \ j = 1,2, \ldots, i-1$, cannot give a PSK-SICG $>g_i$ for the receiver $R_i$  
		\end{enumerate}
		
		\begin{remark}
			Note that the Algorithm \ref{algo_psk} above does not result in a unique mapping of index coded bits to $2^N$- PSK symbols. The mapping will change depending on the choice of $( x_{i_1},x_{i_2}, \ldots, x_{i_{\left|\mathcal{K}_i\right|}} )$ in each step. However, the performance of all the receivers obtained using any such mapping scheme resulting from the algorithm will be the same. Further, if $\eta_{i}= \eta_j$ for some $i \neq j$, depending on the ordering of $\eta_i$ done before starting the algorithm, $R_i$ and $R_j$ may give different performances in terms of probability of error.
		\end{remark}
		
		\subsection{How the Algorithm works}
		For any given realization of $\mathbf{x}= \left( x_1, x_2,\ldots, x_n\right) $, the ML decoder at receiver $R_i$ with $\eta_i < N$ need to consider only  $2^{\eta_i}$ codewords and not the entire $2^N$ possible codewords as explained in the proof of Theorem \ref{th_psk_1}. So the algorithm maps these subset of codewords to PSK signal points to one of the subsets of signal points at the layer $L_{N-\eta_i}$ of Ungerboeck partitioning of the $2^N$-PSK signal set so that these  $2^{\eta_i}$ signal points have a pairwise minimum distance equal to $\Delta_{N-\eta_i}$. An arbitrary mapping cannot ensure this since if any two codewords in this particular subset of $2^{\eta_i}$ codewords are mapped to adjacent points of the $2^N$- PSK signal set, the effective minimum distance seen by the receiver $R_i$ will still be that of $2^N$- PSK. 
		
		Further, since $\Delta_0 < \Delta_1 <  \ldots < \Delta_{N-1}$, the largest pair-wise minimum distance can be obtained by a receiver with the smallest value of $\eta$. Therefore, we order the receivers in the non-decreasing order of their $\eta$ values and map the codewords seen by $R_1$ first, $R_2$ next and so on. Therefore, the largest pair-wise minimum distance and hence the largest PSK-SICG is obtained by $R_1$. 
		
		Consider the index coding problem in Example \ref{ex_psk_1} in Section \ref{sec:Model}.
		Here, $\eta_1=1,\ \eta_2= \eta_3 =2$ and $\eta_i \geq 4, \  i \in \lbrace4,5,6,7\rbrace$.
		While running the Algorithm \ref{algo_psk}, suppose we fix $( x_2, x_3, x_4, x_5, x_6, x_7)= (000000)$,  we get $\mathcal{C}_{1}=\lbrace \lbrace0000\rbrace, \lbrace1000\rbrace \rbrace$. These codewords are mapped to a pair of diametrically opposite 16-PSK symbols, which constitute a subset at the Ungerboeck partition level $L_3$ of the 16-PSK signal set as shown in Fig.  \ref{fig:ex1_map}(a). Then, $\mathcal{C}_{2}$, which results in maximum overlap with $\lbrace \lbrace0000\rbrace, \lbrace1000\rbrace \rbrace$, is  $\lbrace \lbrace0000\rbrace,  \lbrace0100\rbrace, \lbrace1000\rbrace,\lbrace1100\rbrace \rbrace$. We consider $\lbrace0100\rbrace \in \mathcal{C}_2 \setminus \lbrace \lbrace0000 \rbrace, \lbrace1000\rbrace \rbrace$ and map it to a signal point such that the three labeled signal points belong to a subset at level $L_2$. Now we go back to Step 3 with $i=1$ and find $C_1$ which has maximum overlap with the mapped codewords. Now $C_{1} =\lbrace \lbrace0100\rbrace, \lbrace1100\rbrace \rbrace$. Then we map $\lbrace 1100 \rbrace \in \mathcal{C}_1$, which is not already mapped, to that PSK signal point such that $C_{1} =\lbrace \lbrace0100\rbrace, \lbrace1100\rbrace \rbrace$ together constitute a subset at level $L_3$ of the Ungerboeck partitioning. This will result in the mapping  as shown in Fig.  \ref{fig:ex1_map}(b). Continuing in this manner, we finally end up with the mapping shown in Fig. \ref{fig:ex1_map}(c). We see that for such a mapping the $d_{min}^2(R_1)=(2\sqrt{(4)})^2=16$ and  $d_{min}^2(R_2)= d_{min}^2(R_3)=(\sqrt{2}\sqrt{4})^2=8$. 
		
		
		\section{Index Coded Modulation with QAM}
		\label{sec:ICQAM}
		Instead of transmitting $N$ index bits as a point from $2^N$- PSK, we can also transmit the index coded bits as a signal point from $2^N$- QAM signal set, with the average energy of the QAM symbol being equal to the total energy of the $N$ BPSK transmissions. The Algorithm \ref{algo_psk} in Section \ref{sec:PSK_Algo} can be used to map the index coded bits to QAM symbols.
		
		Before starting to run the algorithm to map the index coded bits to $2^N$- QAM symbols, we need to choose an appropriate $2^N$- QAM signal set.
		
		To choose the appropriate QAM signal set, do the following: 
		\begin{itemize}
			\item \textbf{if} $N$ is even,  choose the $2^N$- square QAM with average symbol energy being equal to $N$.
			\item \textbf{else}, take the $2^{N+1}$- square QAM with average symbol energy equal to $N$. Use Ungerboeck set partitioning \cite{TCM} to partition the $2^{N+1}$- QAM signal set into two $2^N$ signal sets. Choose any one of them as the $2^N$- QAM signal set.
		\end{itemize}
		
		After choosing the appropriate signal set, the mapping proceeds in the same way as the mapping of index coded bits to PSK symbols. For the Example \ref{ex_psk_1}, the QAM mapping is shown in Fig. \ref{fig: Map_ex-PSK_QAM}.
		\begin{figure}
			\begin{center}
				\includegraphics[scale= 0.6]{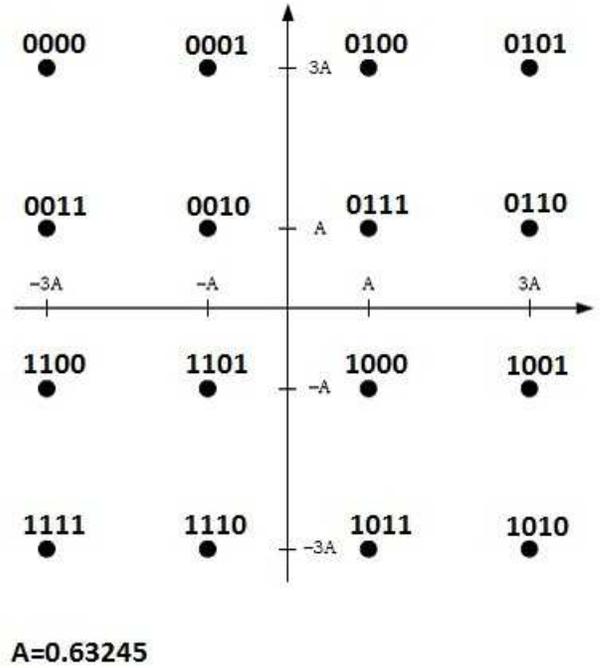}
				\caption{16-QAM mapping for Example \ref{ex_psk_1}}
				\label{fig: Map_ex-PSK_QAM}	
			\end{center}
			~\hrule
			
		\end{figure} 
		The definitions for bandwidth gain, side information coding gain and absolute coding gain are all the same except for the fact that the index coded bits are now transmitted as a QAM signal. Since we transmit QAM signal, we call them QAM bandwidth gain, QAM side information coding gain (QAM-SICG) and QAM absolute coding gain (QAM-ACG) respectively. Further, since the condition for getting SICG depends only on the size of the signal set used, the same set of conditions holds for a receiver to obtain QAM-SICG.
		
		Since for the given index coding problem and for the chosen index code, the index codeword can be transmitted either as a PSK symbol or as a QAM symbol with the conditions for obtaining side information coding gain being same, we need to determine which will result in a better probability of error performance. This is answered in the following theorem.
		
		\begin{theorem}
			\label{th_qam}
			A receiver $R_i$ with $\eta_i \leq 2$ will get better performance when the $N$ index coded bits  are transmitted as a $2^N$- PSK symbol whereas a receiver with $\eta_i > 2$ has better performance when the index coded bits  are transmitted as a $2^N$- QAM symbol.
			\begin{proof} 
				When the $N$ bit index code is transmitted as a signal point from $2^N$- PSK or $2^N$- QAM signal set, the receiver $R_i$ will see an effective signal set of size $2^{\eta_i}$. The side information coding gain for receivers satisfying the condition $\eta_i < N$ comes from mapping the $2^{\eta_i}$ index codewords to signal points on the $2^N$ signal set such that the minimum distance of these $2^{\eta_i}$ signal points is equal to the minimum distance of $2^{\eta_i}$- PSK or QAM and not that of $2^N$- PSK or QAM.
				
				So to prove that for $\eta_i \geq 3$, QAM gives a better error performance, we will show that, for equal average signal energy, $2^{\eta_i}$ points can be mapped to signal points in $2^N$- QAM constellation with a higher minimum distance than to the signal points in $2^N$- PSK. 
				
				The largest possible pair-wise minimum distance that is obtained by any mapping of $2^{\eta}$ points to $2^N$- PSK and QAM signal sets are as follows.
				\begin{align*}
					d_{min-\text{PSK}}(N,\eta) &= 2 \sqrt{N}\sin\left(\frac{\pi}{2^\eta} \right). \\
					d_{min-\text{QAM}}(N,\eta)&=\left\{
					\begin{array}{@{}ll@{}}
						\sqrt{2}^{N-\eta+2}\sqrt{\dfrac{1.5N}{(2^{N}-1)}}, & \text{if}\ N \text{ is even} \\
						\sqrt{2}^{N-\eta+3}\sqrt{\dfrac{1.5N}{(2^{N+1}-1)}}, & \text{otherwise.}
					\end{array}\right.
				\end{align*}  
				
				For sufficiently large values of $N$, 	$d_{min-\text{QAM}}(N,\eta)$ can be approximated for even and odd values of $N$ as $d_{min-\text{QAM}}(N,\eta) \approxeq \sqrt{2}^{2-\eta}\sqrt{1.5N}$. \\
				\textit{Case 1:} For sufficiently large $N$ and $\eta \geq 3$\\
				For $\eta=3, \ \sin\left( \frac{\pi}{2^3}\right) = 0.3827$ and $\frac{\pi}{2^3} = 0.3927$. Therefore for all $\eta \geq 3$, we take $\sin\left( \frac{\pi}{2^3}\right)\approxeq \frac{\pi}{2^\eta}$.
				Therefore, we have 
				\begin{align*}
					d_{min-\text{PSK}}(N,\eta) & \approxeq 2 \sqrt{N}\left(\frac{\pi}{2^\eta} \right)\\ d_{min-\text{QAM}}(N,\eta) & \approxeq \sqrt{2}^{2-\eta}\sqrt{1.5N}.
				\end{align*} 
				We see that $\frac{d_{min-\text{QAM}}(N,\eta)}{d_{min-\text{PSK}}(N,\eta)}= \left( \frac{\sqrt{1.5}}{\pi}\right) \sqrt{2}^\eta \geq 1, \forall \ \eta \geq 3.$
				Therefore QAM gives a better performance than PSK if $\eta \geq 3$ for sufficiently large.\\
				\textit{Case 2:} For sufficiently large $N$ and $\eta = 1,2$.\\
				With $\eta=1$, we have
				\begin{align*}
					d_{min-\text{PSK}}(N,1) & = 2 \sqrt{N}\sin\left(\frac{\pi}{2^\eta} \right)  =2\sqrt{N} \\ d_{min-\text{QAM}}(N,1) & \approxeq \sqrt{2}^{2-\eta}\sqrt{1.5N} = \sqrt{3N}.  
				\end{align*}  
				Clearly, $d_{min-\text{PSK}}(N,1) > d_{min-\text{QAM}}(N,1)$. Therefore, PSK has a better performance.
				
				Similarly with $\eta=2$, we have $d_{min-\text{PSK}}(N,2) = \sqrt{2N}$ and $d_{min-\text{QAM}}(N,2) = \sqrt{1.5N}.$ Again, PSK performs better than QAM.\\
			\end{proof}
		\end{theorem}
		
		\begin{figure*}
			\includegraphics[scale=0.4]{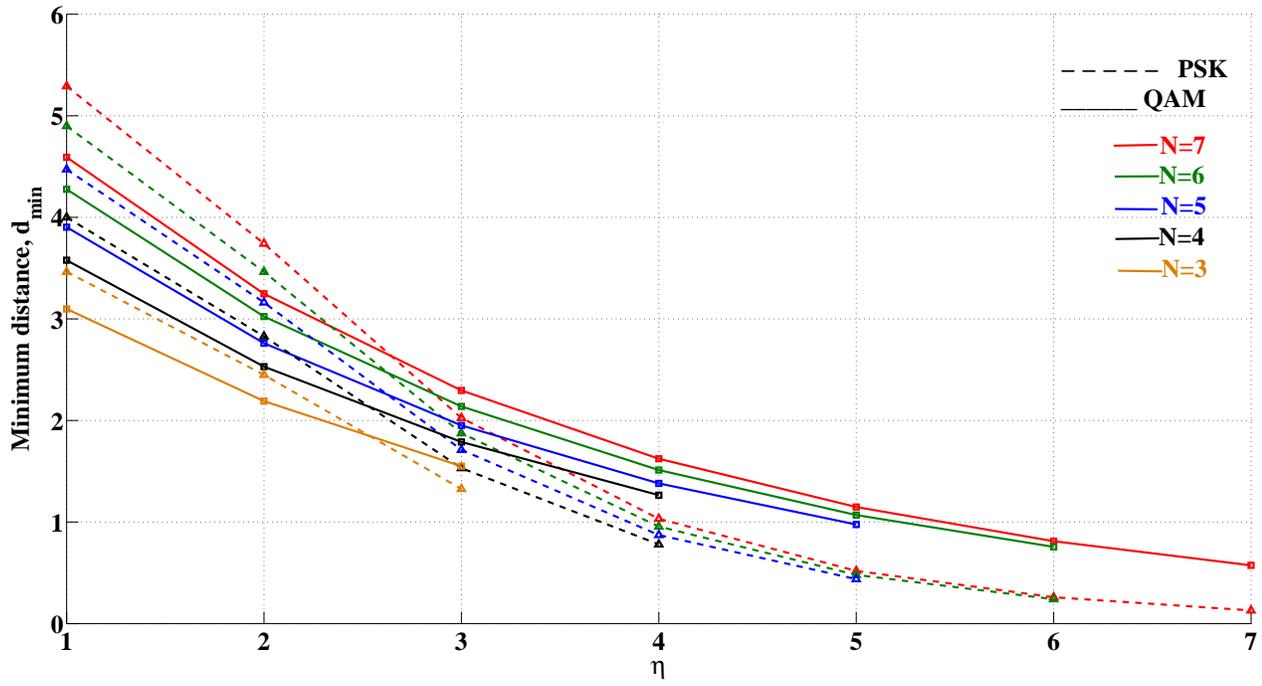}
			\caption{Minimum distance of PSK and QAM for different values of $N$ and $\eta$}
			\label{fig: min_dis_N_eta}
			~\hrule
			
		\end{figure*}
		
		We have also given a plot validating  the result for $N=3,4,5,6$ and $7$ in Fig. \ref{fig: min_dis_N_eta}. Hence we see that for receivers with less amount of side information, i.e., receivers which see effective signal sets with eight points or more, transmitting the index codeword as a QAM symbol will result in better probability of error performance.
		
		
		\section{Simulation Results}
		\label{sec:simu}
		
		Simulation results for the Example \ref{ex_psk_1} is shown in Fig. \ref{fig:ex1_sim}. We see that the probability of message error plots corresponding to $R_{1}$ is well to the left of the plots of $R_{2}$ and $R_{3}$, which themselves are far to the left of other receivers as $R_{1},\ R_{2},\ R_{3}$ get PSK-SICG as defined in Section \ref{sec:Model}. Since $\left|S_{1}\right|>\left|S_{2}\right|=\left|S_{3}\right|, R_{1}$ gets the highest PSK-SICG. Further, since $\mathcal{K}_{4}, \ \mathcal{K}_{5},\ \mathcal{K}_{6}$ and $\mathcal{K}_{7}$ does not satisfy any of the two conditions required, they do not get PSK-SICG. The performance improvement gained by $R_1, R_2$ and $R_3$ over 4-fold BPSK index code transmission can also be observed.
		
		\begin{figure*}
			
			\includegraphics[scale=0.4]{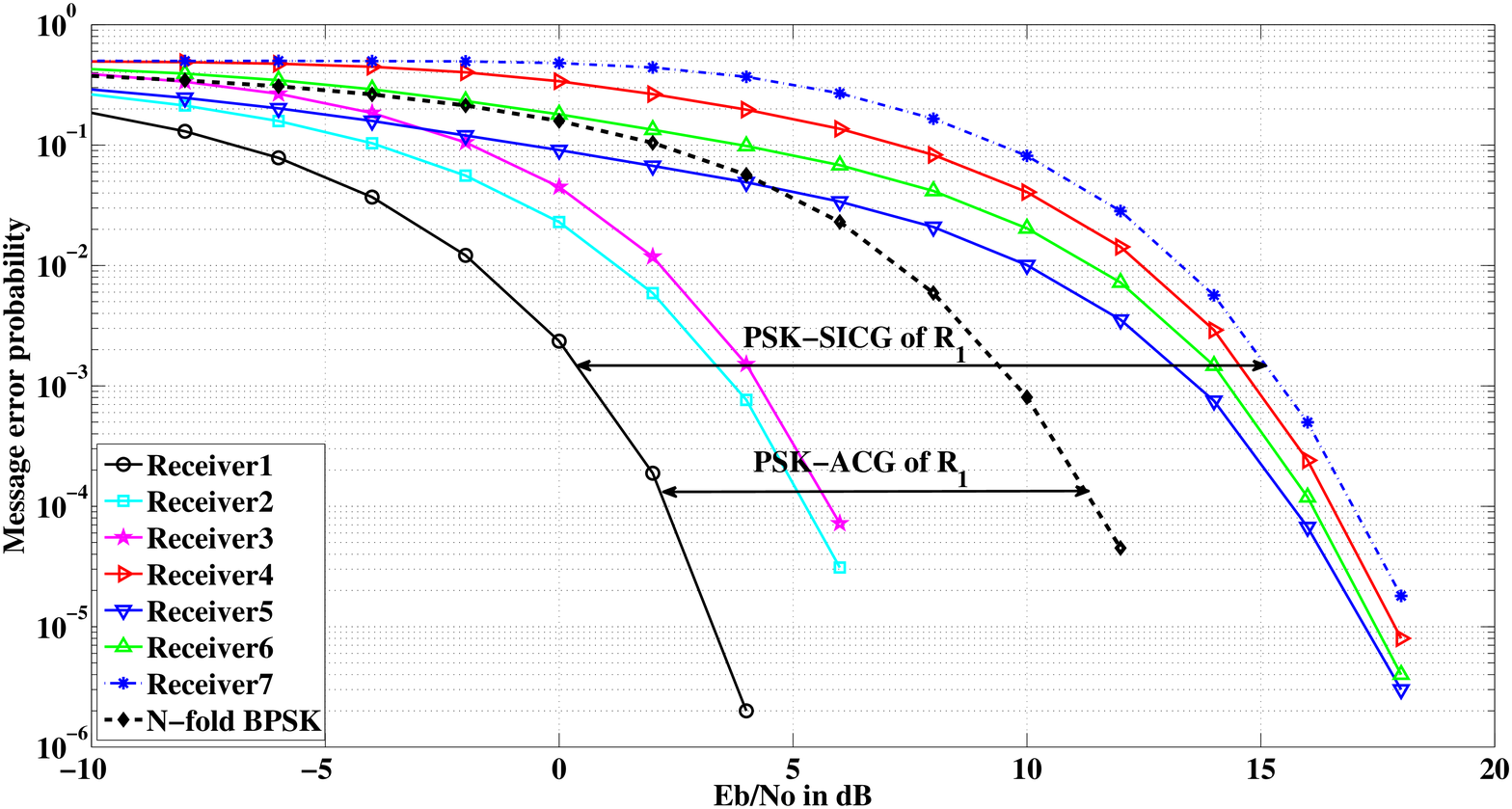}
			\caption{Simulation results for Example \ref{ex_psk_1}.}
			\label{fig:ex1_sim}
		\end{figure*}
		
		
		From the probability of message error plot, though it would seem that the receivers $R_{4}, R_{5}, R_{6}$ and $R_{7}$ lose out in probability of message error performance to the 4-fold BPSK scheme, they are merely trading off coding gain for bandwidth gain as where the 4-fold BPSK scheme for this example uses 4 real dimensions, the proposed scheme only uses 1 complex dimension, i.e., 2 real dimensions. Hence the receivers $R_{4},\ R_{5},\ R_{6}$ and $R_{7}$ get PSK bandwidth gain even though they do not get PSK-ACG whereas $R_{1}$, $R_{2}$ and $R_{3}$ get both PSK bandwidth gain and PSK-ACG. The amount of PSK-SICG, PSK bandwidth gain and PSK-ACG that each receiver gets is summarized in TABLE \ref{Table1}.
		
		{\footnotesize
			\begin{table}[h]
				\renewcommand{\arraystretch}{1.25}
				\begin{center}
					
					\begin{tabular}{|m{2cm}|c|c|c|c|c|c|c|}
						\hline
						Parameter & $R_{1}$ &  $R_{2}$ & $R_{4}$ & $R_{5}$ & $R_{6}$ & $R_{7}$ \\
						\hline 
						$d_{min_{PSK}}^2$ & 16 &  8 & 0.61 & 0.61 & 0.61 & 0.61  \\ 
						
						$d_{min_{binary}}^2$ & 4  & 4 & 4 & 4 & 4 & 4 \\ 
						
						PSK bandwidth gain & 2  & 2 & 2 & 2 & 2 & 2 \\
						
						PSK-SICG (in dB) & 14.19  & 11.19 & 0 & 0 & 0 & 0 \\
						
						PSK-ACG (in dB) & 6.02  & 3.01 & -8.16 & -8.16 & -8.16 & -8.16\\
						
						\hline
						
					\end{tabular}
					
					\caption \small { Table showing  PSK-SICG, PSK bandwidth gain and PSK-ACG for different receivers in Example \ref{ex_psk_1}. $R_3$ has same values as $R_2$}
					
					\label{Table1}	
					
				\end{center}
			\end{table}
		}
		
		\begin{figure}
			\begin{center}
				\includegraphics[scale=0.42]{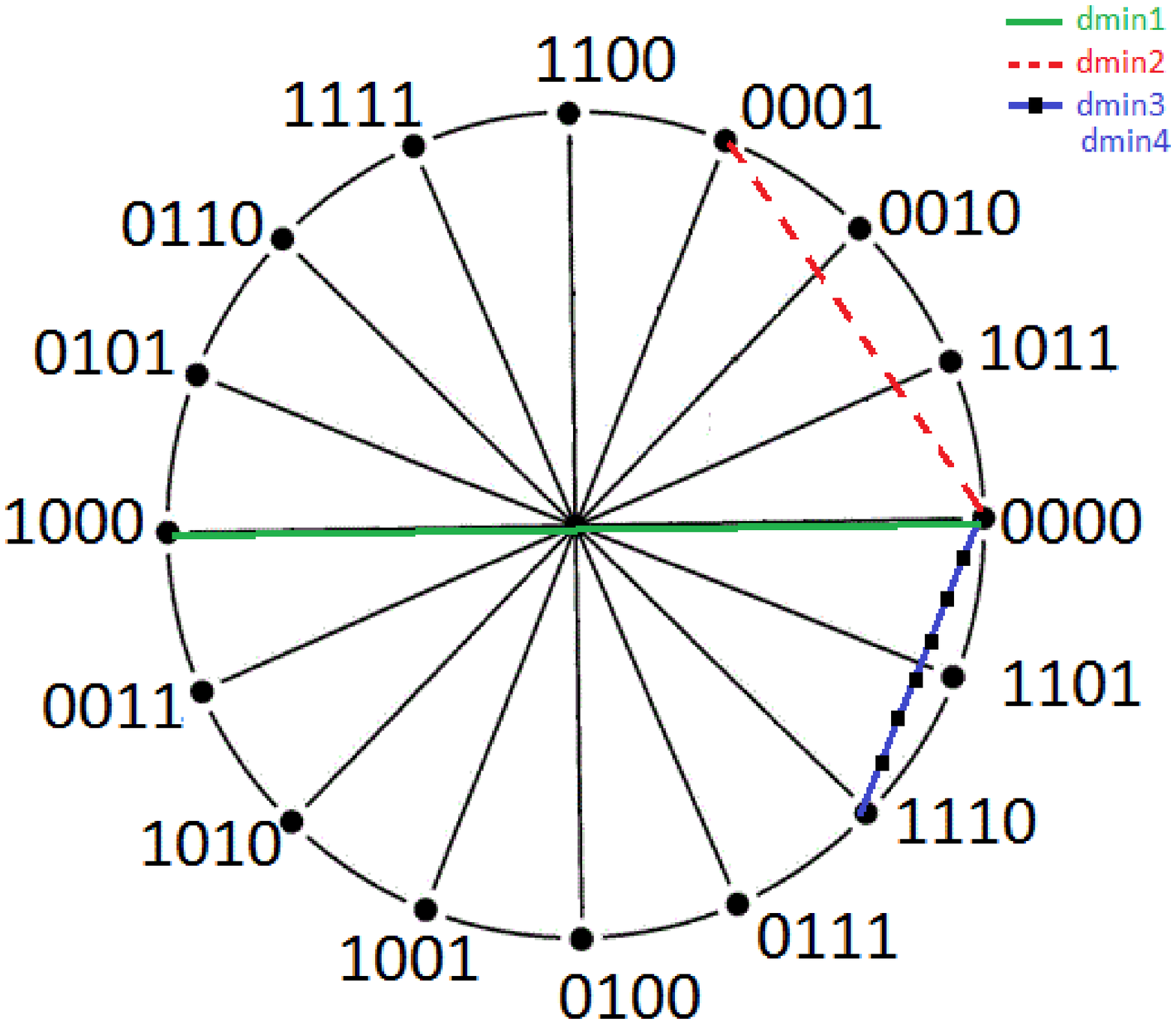}
				\caption{16-PSK Mapping for Example \ref{ex3}.}
				\label{fig6}
			\end{center}
			~\hrule
		\end{figure}
		\begin{figure*}
			\includegraphics[scale=0.4]{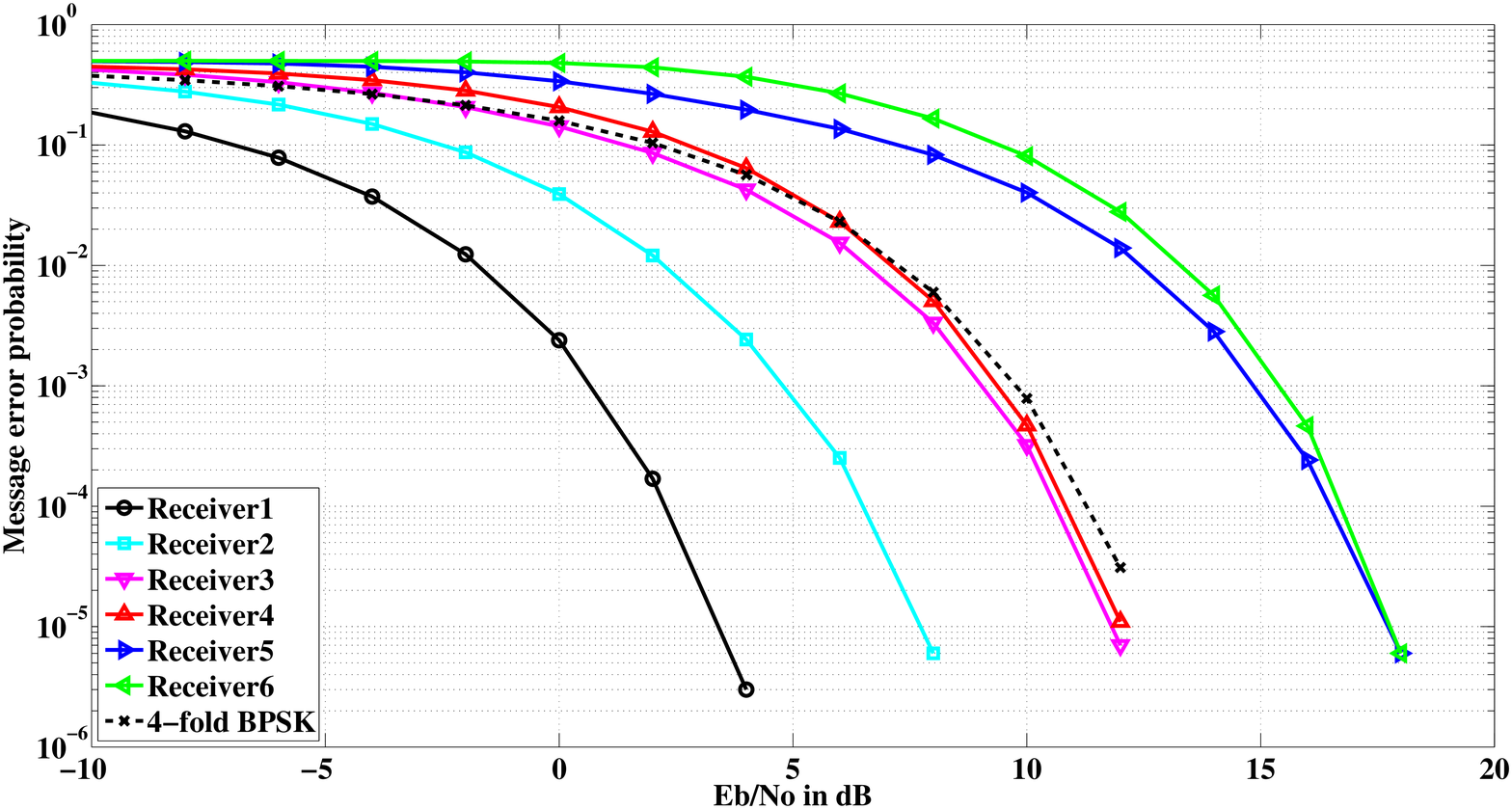}
			\caption{Simulation results for Example \ref{ex3}.}
			\label{fig5}
		\end{figure*}

		Now consider Example \ref{ex3}. Here, suppose we fix  $( x_2, x_3, x_4, x_5, x_6)= (00000)$, we get  $\mathcal{C}_{1}=\lbrace \lbrace0000\rbrace, \lbrace1000\rbrace \rbrace$. After mapping these codewords to a subset at level $L_3$ of the Ungerboeck partition of the 16-PSK signal set, a subset of $\mathcal{C}$ which results in maximum overlap with already mapped codewords is  $\mathcal{C}_{2} = \lbrace \lbrace0000\rbrace,  \lbrace0001\rbrace, \lbrace0100\rbrace,\lbrace0101\rbrace \rbrace$.	We see that $\mathcal{C}_{1}\not \subseteq \mathcal{C}_{2}$, so all the codewords in $\mathcal{C}_{2}$ cannot be mapped to the same 4-point subset in the level $L_2$ without disturbing the mapping of codewords of $\mathcal{C}_{1}$ already done. So we try to map them in such a way that the minimum distance,  $d_{min}(R_2) \geq d_{min}$ of  8-PSK. The algorithm gives a mapping which gives the best possible $d_{min}(R_2)$ keeping  $d_{min}(R_1) = d_{min}$ of 2-PSK. This mapping is shown in Fig. \ref{fig6}.
		
		Simulation results for this example is shown in Fig. \ref{fig5}. The receivers $R_{1}, R_{2}, R_{3}$ and $R_{4}$ get PSK-SICG. We see that the probability of message error plots corresponding to the 4-fold BPSK binary transmission scheme lies near $R_{3}$ and $R_{4}$ showing better performances for receivers $R_{1}$ and $R_{2}$. Thus receivers $R_{1}$ and $R_{2}$ get PSK-ACG as well as PSK bandwidth gain over the 4-fold BPSK scheme, $R_{3}$ and $R_{4}$ get the same performance as 4-fold BPSK with additional bandwidth gain and $R_{5}$ and $R_{6}$ trade off bandwidth gain for coding gain. The amount of PSK-SICG, PSK bandwidth gain and PSK-ACG that each receiver gets is summarized in TABLE \ref{Table3}.
		\begin{table}[h]
			\renewcommand{\arraystretch}{1.25}
			\begin{center}
				
				\begin{tabular}{|c|c|c|c|c|c|c|}
					\hline
					Parameter & $R_{1}$ & $R_{2}$ & $R_{3}$ & $R_{4}$ & $R_{5}$ & $R_{6}$  \\
					\hline 
					$d_{min_{PSK}}^2$ & 16 & 4.94 & 2.34 & 2.34 & 0.61 & 0.61   \\ 
					
					$d_{min_{binary}}^2$ & 4 & 4 & 4 & 4 & 4 & 4 \\ 
					
					PSK bandwidth gain & 2 & 2 & 2 & 2 & 2 & 2  \\
					
					PSK-SICG (in dB) & 14.19 & 9.08 & 5.84 & 5.84 & 0 & 0  \\
					
					PSK-ACG (in dB) & 6.02 & 0.92 & -2.33 & -2.33 & -8.16 & -8.16 \\
					
					\hline
					
				\end{tabular}
				\caption \small { Table showing  PSK-SICG, PSK bandwidth gain and PSK-ACG for different receivers in Example \ref{ex3}.}
				\label{Table3}	
			\end{center}
		\end{table}
		\begin{remark}
			Even though the minimum distance for the 4-fold BPSK transmissions is better than $d_{min}(R_3)$ and $d_{min}(R_4)$, as seen from TABLE \ref{Table3}, the probability of error plot for the 4-fold BPSK lies slightly to the right of the error plots for $R_3$ and $R_4$. This is because since the 4-fold BPSK scheme takes 2 times the bandwidth used by the 16-PSK scheme, the noise power = $N_o$(Bandwidth), where $N_o$ is the noise power spectral density, is 2 times more for the 4-fold BPSK. Therefore, the signal to noise power ratio for the 4-fold BPSK scheme is 2 times less than that for 16-PSK scheme, even though the transmitted signal power is the same for both the schemes. 
		\end{remark}
		\begin{figure*}
			\begin{center}	
				\includegraphics[scale=0.4]{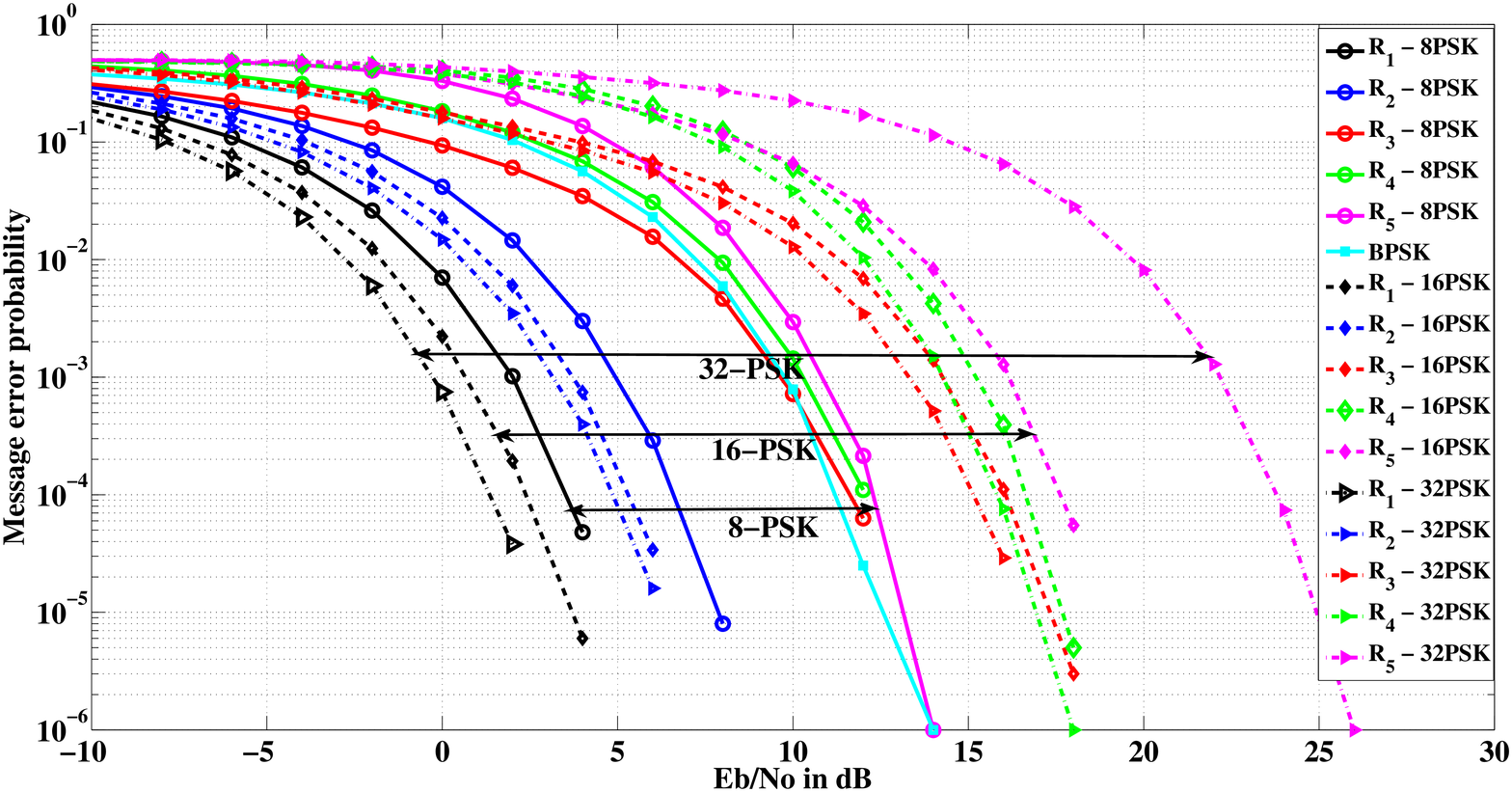}
				\caption{Simulation results for Example \ref{ex_N_to_n1}.}
				\label{fig12}
			\end{center}	
			~\hrule
			
		\end{figure*}
		The following example demonstrates that if $\eta_{i}= \eta_j$ for some $i \neq j$, depending on the ordering of $\eta_i$ done before starting the algorithm, the mapping changes and hence the probability of error performances of $R_i$ and $R_j$ can change.
		\begin{example}
			\label{ex4}
			Let $m=n=6$ and the demanded messages be  $\mathcal{W}_i = x_{i}, \ \forall \ i\in \lbrace 1, 2, \ldots, 6 \rbrace $. The side information possessed by various receivers are
			$\mathcal{K}_1 =\left\{2, 4, 5, 6\right\},\ \mathcal{K}_2=\left\{1, 3, 4, 5\right\},\ \mathcal{K}_3=\left\{2,4\right\},\ \mathcal{K}_4=\left\{1,3\right\},\ \mathcal{K}_5=\left\{2\right\},\ \text{and} \ \mathcal{K}_6=\left\{1\right\}$.\\
			For this problem, the minrank $N$=3. 
			An optimal linear index code is given by the encoding matrix,
			
			{\small
				\begin{center}
					
					$L = \left[\begin{array}{ccc}
					1 & 0 & 0 \\
					0 & 1 & 0 \\
					0 & 0 & 1 \\
					0 & 0 & 1 \\
					0 & 1 & 0 \\
					1 & 0 & 0 
					\end{array}\right]$. \\
				\end{center}
			}	
			
			$ \mbox{ Here~~~~~~~~~~~~~~~~~~~~~~} y_{1}=x_{1}+x_{6};~~~~~~~~~~	y_{2}=x_{2}+x_{5}; ~~~~~~~~~~	y_{3}=x_{3}+x_{4}.$
			
			\begin{figure}[h]
				\begin{center}
					\includegraphics[scale=0.39]{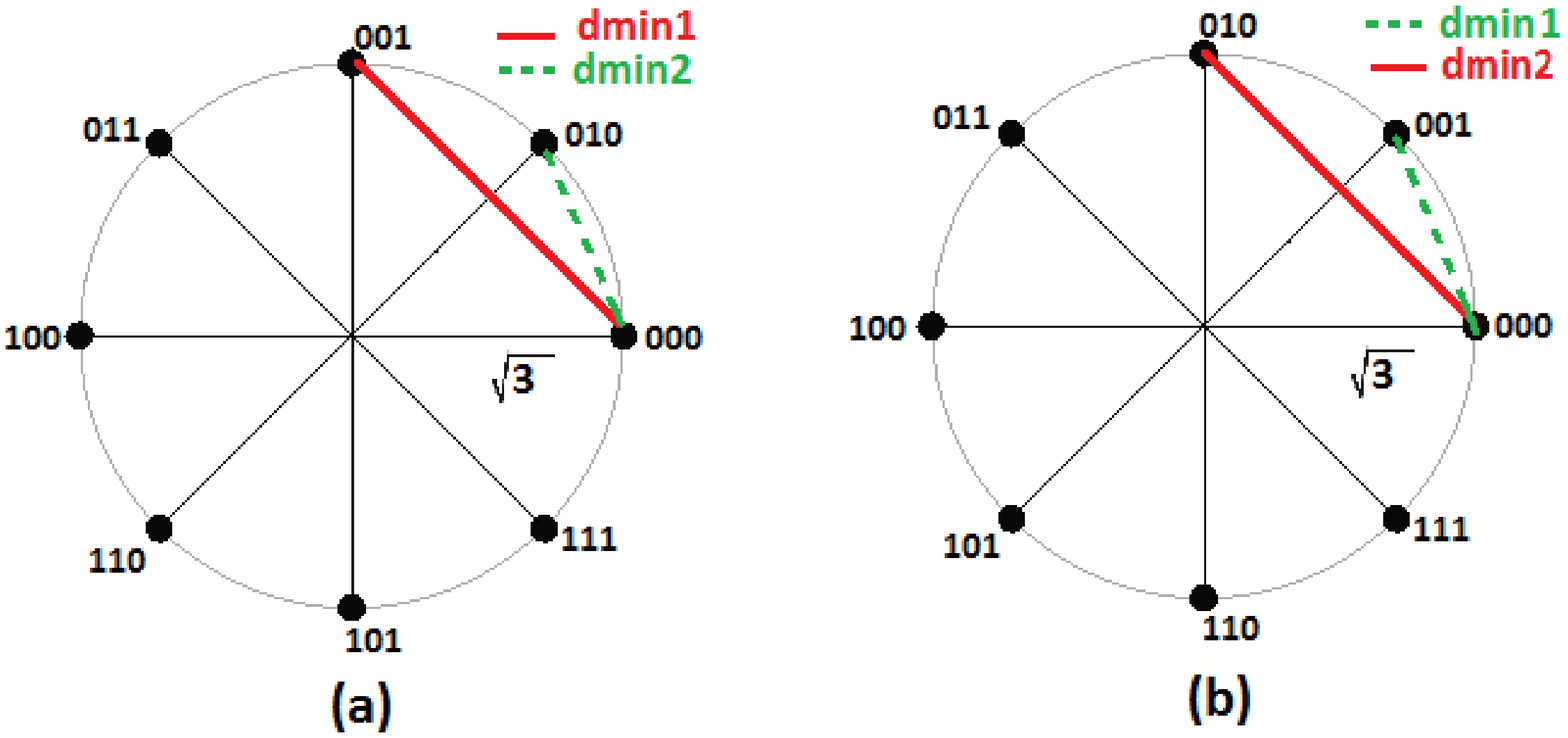}
					\caption{8-PSK Mappings for the 2 cases in Example \ref{ex4}.}
					\label{fig7}
				\end{center}
				~\hrule
			\end{figure}
			\begin{figure*}
				\includegraphics[scale=0.4]{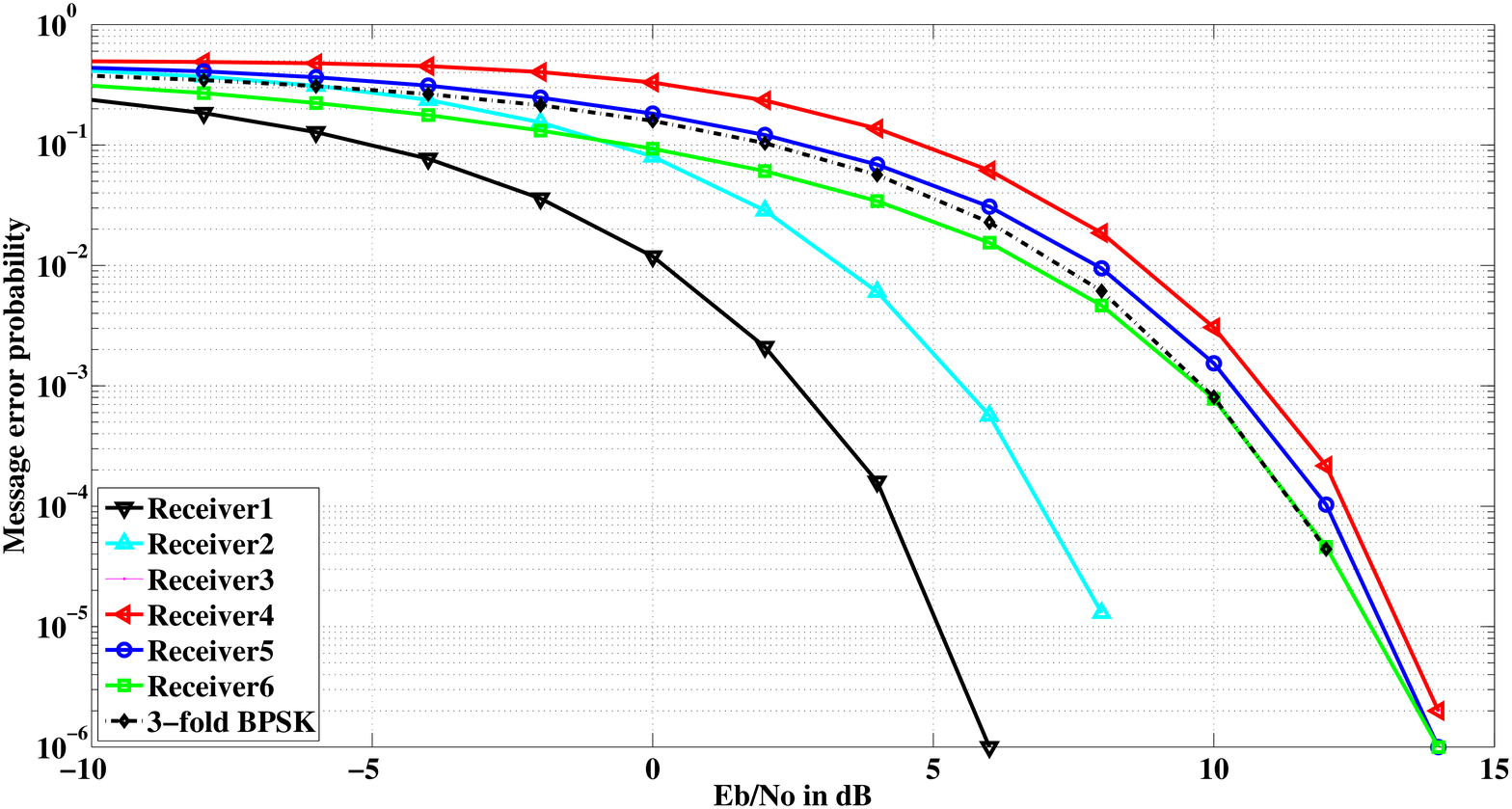}
				\caption{Simulation results for Example \ref{ex4}.}
				\label{fig9}
			\end{figure*}
			We see that $\left|\mathcal{K}_1\right| = \left|\mathcal{K}_2\right|$ and $\left|S_{1}\right| = \left|S_2\right|,\ \therefore \eta_1 = \eta_2$. Then, we can choose to prioritize $R_{1}$ or $R_{2}$ depending on the requirement. If we choose $R_{1}$, the resulting mapping is shown in Fig. \ref{fig7}(a) and if we choose $R_{2}$, then the mapping is shown in Fig. \ref{fig7}(b). Simulation results for this example with the mapping in Fig. \ref{fig7}(a) is shown in Fig. \ref{fig9}, where we can see that $R_{1}$ outperforms the other receivers. $R_{1}$ and $R_{2}$ get PSK-SICG as expected. They also get PSK-ACG. The other receivers have the same performance as the 3-fold BPSK scheme. All 6 receivers get PSK bandwidth gain. The amount of PSK-SICG, PSK bandwidth gain and PSK-ACG that each receiver gets is summarized in TABLE \ref{Table4}.	
			
			{\small	
				\begin{table}[h]
					\renewcommand{\arraystretch}{1.25}
					\begin{center}
						
						\begin{tabular}{|c|c|c|c|c|c|c|}
							\hline
							Parameter & $R_{1}$ & $R_{2}$ & $R_{3}$ & $R_{4}$ & $R_{5}$ & $R_{6}$  \\
							\hline 
							$d_{min_{PSK}}^2$ & 6 & 1.76 & 1.76 & 1.76 & 1.76 & 1.76   \\ 
							
							$d_{min_{binary}}^2$ & 4 & 4 & 4 & 4 & 4 & 4 \\ 
							
							PSK bandwidth gain & 1.5 & 1.5 & 1.5 & 1.5 & 1.5 & 1.5  \\
							
							PSK-SICG (in dB) & 5.33 & 0 & 0 & 0 & 0 & 0  \\
							
							PSK-ACG (in dB) & 1.77 & -3.56 & -3.56 & -3.56 & -3.56 & -3.56 \\
							
							\hline
							
						\end{tabular}
						
						\caption \small { Table showing  PSK-SICG, PSK bandwidth gain and PSK-ACG for different receivers for case (a) in Example \ref{ex4}.}
						
						\label{Table4}	
						
					\end{center}
				\end{table}
			}
		\end{example}
		
		Here, even though $d_{min}(R_2)=d_{min}(R_3)=d_{min}(R_4)=d_{min}(R_5)=d_{min}(R_6)$, the probability of error plot of $R_2$ is well to the left of the error plots of $R_3$, $R_4$, $R_5$ and $R_6$. This is because the distance distribution seen by $R_2$ is different from the distance distribution seen by the other receivers, as shown in TABLE \ref{Table5}, where, $d_{min_1}$ gives the minimum pairwise distance, $d_{min_2}$ gives the second least pairwise distance and so on.
		
		{\small
			\begin{table}
				\renewcommand{\arraystretch}{1.25}
				\begin{center}
					
					\begin{tabular}{|c|c|c|c|c|c|c|}
						\hline
						Parameter & $R_1$ & $R_2$ & $R_3$ & $R_4$ & $R_5$ & $R_6$ \\
						\hline
						Effective signal set seen & 4 pt & 4 pt & 8 pt & 8 pt & 8 pt & 8 pt \\
						$d_{min_1}$ & 6 & 1.76 & 1.76 & 1.76 & 1.76 & 1.76 \\
						No. of pairs & 4 & 4 & 8 & 8 & 8 & 8\\
						$d_{min_2}$ & 12 & 10.24 & 6 & 6 & 6 & 6 \\
						No. of pairs & 2 & 2 & 8 & 8 & 8 & 8 \\
						$d_{min_3}$ & -- & 12 & 10.24 & 10.24 & 10.24 & 10.24\\
						No. of pairs & 0 & 2 & 8 & 8 & 8 & 8 \\
						$d_{min_4}$ & -- & -- & 12 & 12 & 12 & 12 \\
						No. of pairs & 0 & 0 & 4 & 4 & 4 & 4 \\
						\hline			
					\end{tabular}
					\caption \small {Table showing the pair-wise distance distribution for the receivers in Example \ref{ex4}.}
					\label{Table5}
				\end{center}	
			\end{table}
		}
		\subsection{$2^N$-PSK to $2^n$-PSK}
		
		The simulation results for the Example \ref{ex_N_to_n1} is shown in Fig. \ref{fig12}. We can see that the best performing receiver's, i.e., $R_1$'s  performance improves as we go from $N$ to $n$. The minimum distance seen by different receivers for the 3 cases considered, namely, 8-PSK, 16-PSK and 32-PSK, are listed in TABLE \ref{Table7}.

		\begin{table}[h]
			\renewcommand{\arraystretch}{1.25}
			\begin{center}
				
				\begin{tabular}{|c|c|c|c|c|c|}
					\hline
					Parameter & $R_{1}$ & $R_{2}$ & $R_{3}$ & $R_{4}$ & $R_{5}$ \\
					\hline 
					$d_{min_{\ 8-PSK}}^2$ & 12 & 6 & 1.76 & 1.76 & 1.76 \\ 
					
					$d_{min_{\ 16-PSK}}^2$ & 16 & 8 & 0.61 & 0.61 & 0.61   \\ 
					
					$d_{min_{\ 32-PSK}}^2$ & 20 & 8.05 & 0.76 & 0.76 & 0.19  \\ 
					
					$d_{min_{binary}}^2$ & 4 & 4 & 4 & 4 & 4 \\ 
					
					\hline
					
				\end{tabular}
				
				\caption \small { Table showing  the minimum distances seen by different receivers for 8-PSK, 16-PSK and 32-PSK in  Example \ref{ex_N_to_n1}.}
				
				\label{Table7}	
				
			\end{center}
		\end{table}
	\end{example}
	This example satisfies the condition in Lemma \ref{lemm:N_to_n} and hence the difference in performance between $R_1$ and $R_5$ increases monotonically with the length of the index code used. However, as stated in the Remark \ref{rem:N_to_n}, when the receiver with the worst probability of error performance knows at least one message a priori, the difference between the performances of the best and worst receiver need not increase monotonically. This is illustrated in the following example.	
	
	
	\begin{example}
		\label{ex5}
		Let $m=n=$ 4 and $\mathcal{W}_i = x_{i},\ \forall \ i\in \lbrace 1, 2, \ldots, 4 \rbrace $, with the side information sets being  
		$\mathcal{K}_1 =\left\{2, 3, 4\right\},\ \mathcal{K}_2=\left\{1, 3\right\},\ \mathcal{K}_3=\left\{1,4\right\}$ and  $\mathcal{K}_4=\left\{2\right\}$. \\
		For this problem, the minrank evaluates to $N$=2. 
		An optimal linear index code is given by the encoding matrix,\\
		
		\begin{center}
			
			$L_1 = \left[\begin{array}{ccc}
			1 & 0  \\
			1 & 1  \\
			1 & 0  \\
			0 & 1  
			
			\end{array}\right]$.\\
		\end{center}
		
		The corresponding 4-PSK mapping is given in Fig. \ref{fig8}(a). 
		
		\begin{figure*}
			\includegraphics[scale=0.4]{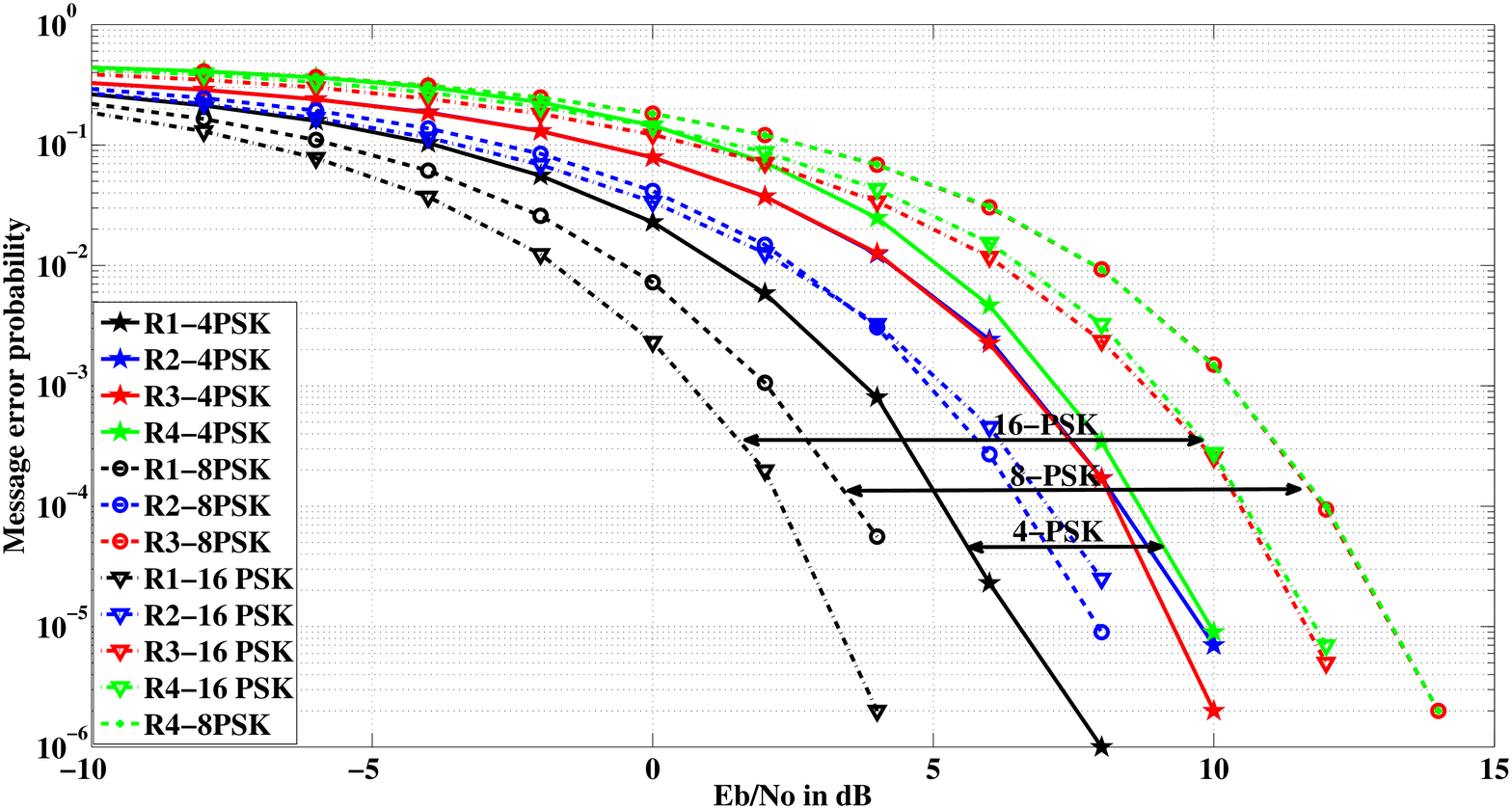}
			\caption{Simulation results for Example \ref{ex5}.}
			\label{fig10}
			~\hrule
		\end{figure*}
		\begin{figure*}[h]
			\includegraphics[scale=0.4]{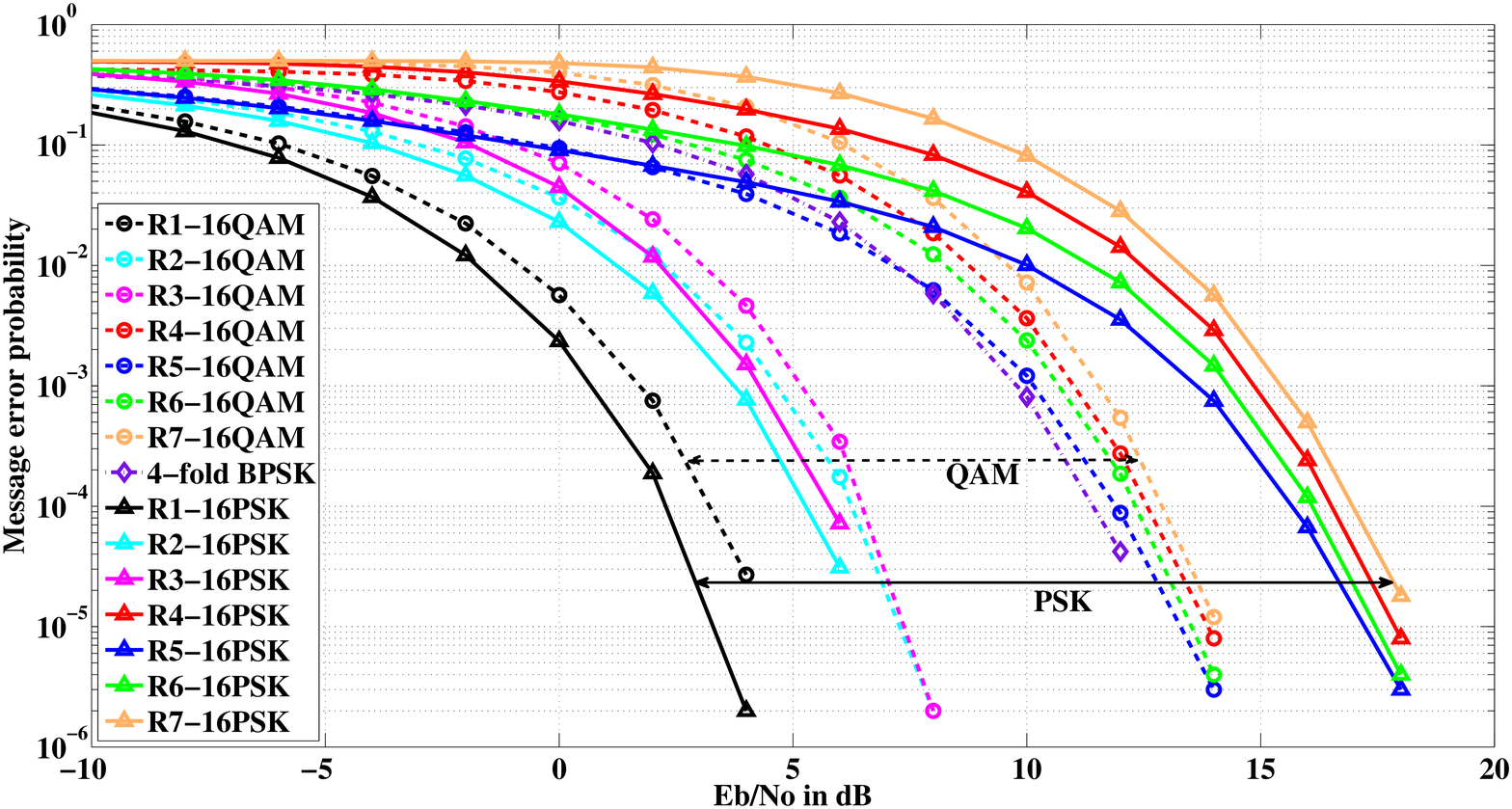}
			\caption{Simulation result comparing the performance of 16-PSK and 16-QAM for Example \ref{ex_psk_1}.}
			\label{fig: sim_ex-PSK_QAM}
			~\hrule
		\end{figure*}
		\begin{figure}[h]
			\begin{center}
				\includegraphics[scale=0.4]{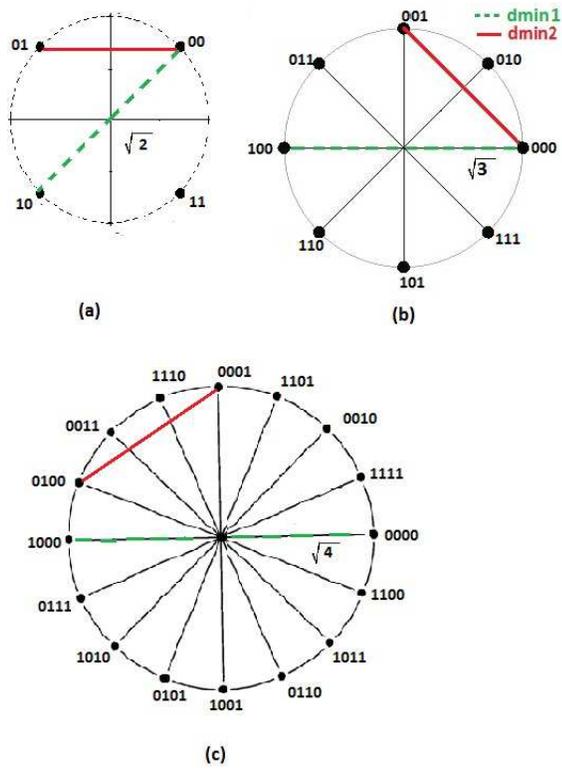}
				\caption{4-PSK, 8-PSK and 16-PSK Mappings for  Example \ref{ex5}.}
				\label{fig8}
			\end{center}
			~\hrule
		\end{figure}
		
		Now assuming  that we did not know the minrank for the above problem and chose $N=3$. Then an encoding matrix is, 
		\begin{center}
			
			$L_2 = \left[\begin{array}{ccc}
			1 & 0 & 0 \\
			1 & 0 & 0 \\
			0 & 1 & 0 \\
			0 & 0 & 1 \\
			
			\end{array}\right]$, \\
		\end{center}
		
		
		and an 8-PSK mapping which gives the best possible PSK-SICGs for the different receivers is shown in Fig. \ref{fig8}(b).

		Now, compare the above two cases with the case where the 4 messages are transmitted as they are, i.e., $ \left[y_1 \ y_2\ y_3\ y_4\right]=\left[x_1\ x_2\ x_3\ x_4\right]$. A 16-PSK mapping which gives the maximum possible PSK-SICG is shown in Fig. \ref{fig8}(c).
		
		From the simulation results shown in Fig \ref{fig10}, we see that the performance of the best receiver, i.e., $R_1$, improves as we go from $N$ to $n$. However, the gap between the best performing receiver and worst performing receiver widens as we go from N to n. The reason for the difference in performance seen by different receivers is that they see different minimum distances, which is summarized in TABLE \ref{Table6}, for 4-PSK, 8-PSK and 16-PSK.
		
		\begin{table}[h]
			\renewcommand{\arraystretch}{1.25}
			\begin{center}
				\begin{tabular}{|c|c|c|c|c|}
					\hline
					Parameter & $R_{1}$ & $R_{2}$ & $R_{3}$ & $R_{4}$  \\
					\hline 
					$d_{min_{\ 4-PSK}}^2$ & 8 & 4 & 4 & 4   \\ 
					
					$d_{min_{\ 8-PSK}}^2$ & 12 & 6 & 1.76 & 1.76   \\ 
					$d_{min_{\ 16-PSK}}^2$ & 16 & 4.94 & 2.34 & 2.34   \\ 
					$d_{min_{binary}}^2$ & 4 & 4 & 4 & 4  \\ 
					
					\hline
					
				\end{tabular}
				
				\caption \small { Table showing  the minimum distances seen by different receivers for 4-PSK, 8-PSK and 16-PSK in  Example \ref{ex5}.}
				
				\label{Table6}	
				
			\end{center}
		\end{table}
		
	\end{example}
	Here we see that the difference in performance between the best and worst receiver is not monotonically widening with the length of the index code employed.
	\subsection{Comparison between PSK and QAM}
	
	For the Example \ref{ex_psk_1}, the  plot comparing the performances of PSK and QAM is shown in Fig.  \ref{fig: sim_ex-PSK_QAM}.

	\begin{table}[h]
		\renewcommand{\arraystretch}{1.25}
		\begin{center}
			
			\begin{tabular}{|c|c|c|c|c|c|c|c|}
				\hline
				Parameter & $R_{1}$ &  $R_{2}$ & $R_{4}$ & $R_{5}$ & $R_{6}$ & $R_{7}$ \\
				\hline 
				
				$d_{min}^2 - 16-QAM$ & 12.8 & 6.4 & 1.6 & 1.6 & 1.6 & 1.6 \\ 
				
				$d_{min}^2 - 16-PSK$ & 16 &  8 & 0.61 & 0.61 & 0.61 & 0.61  \\ 
				
				$d_{min}^2 - binary$ & 4  & 4 & 4 & 4 & 4 & 4 \\ 				
				\hline
				
			\end{tabular}
			
			\caption \small { Table showing  minimum distance seen by different receivers while using 16-QAM and 16-PSK in Example \ref{ex_psk_1}. $R_3$ has same values as $R_2$.}
			
			\label{Table-ex_PSK_QAM}	
			
		\end{center}
	\end{table}
	
	We can see that while $R_1, \ R_2 \ \text{and} \ R_3$ performs better while the index coded bits are transmitted as a PSK signal, the other receivers have better performance when a QAM symbol is transmitted.  This is because of the difference in the minimum distances seen by the different receivers as summarized in TABLE \ref{Table-ex_PSK_QAM}. This observation agrees with Theorem \ref{th_qam}.

	
	\section{Conclusion}
	\label{sec:conc}
	The mapping and 2-D transmission scheme proposed in this paper is applicable to any index coding problem setting. In a practical scenario, we can use this mapping scheme to prioritize those customers who are willing to pay more, provided their side information satisfies the condition mentioned in Section \ref{sec:Model}. Further, the mapping scheme depends on the index code, i.e., the encoding matrix, $L$, chosen, since $L$ determines $\left|S_{i}\right|, \forall \ i \in \lbrace1, 2, \ldots, m\rbrace$. So we can even choose an $L$ matrix such that it favors our chosen customer, provided $L$ satisfies the condition that all users use the minimum possible number of binary transmissions to decode their required messages. Further, if we are interested only in giving the best possible performance to a chosen customer who has large amount of side information and not in giving the best possible performance to every receiver, then using a $2^n$- PSK/QAM would be a better strategy. The mapping and 2-D transmission scheme introduced in this paper are also applicable to index coding over fading channels which was considered in \cite{OLIC}.
	\section{Acknowledgment}
	This work was supported partly by  the Science and Engineering Research Board (SERB) of Department of Science and Technology (DST), Government of India, through  J.C. Bose National Fellowship to B. Sundar Rajan.

\end{document}